\documentclass[pdflatex,sn-mathphys-num]{sn-jnl}

\usepackage{graphicx}%
\usepackage{multirow}%
\usepackage{amsmath,amssymb,amsfonts}%
\usepackage{amsthm}%
\usepackage{mathrsfs}%
\usepackage[title]{appendix}%
\usepackage{xcolor}%
\usepackage{textcomp}%
\usepackage{manyfoot}%
\usepackage{booktabs}%
\usepackage{algorithm}%
\usepackage{algorithmicx}%
\usepackage{algpseudocode}%
\usepackage{listings}%
\usepackage{siunitx}%
\usepackage{isotope}%
\usepackage{bm}%
\let\orcidlogo\relax
\usepackage{orcidlink}
\DeclareSIUnit\clight{\ensuremath{c}}%
\DeclareSIUnit\nanobarn{nb}%
\newcounter{subfignum}[figure]%
\newlength{\subfiglabelraise}%
\renewcommand{\thesubfignum}{\the\numexpr\value{figure}+1\relax\alph{subfignum}}%
\newcommand{\subfiglabel}[2][]{%
  \refstepcounter{subfignum}%
  \raisebox{\subfiglabelraise}[\height][\depth]{%
    {\figurecaptionfont(\alph{subfignum})%
      \if\relax\detokenize{#1}\relax\else\hskip.7em#1\fi}}%
  \label{#2}}%

\theoremstyle{thmstyleone}%
\theoremstyle{thmstyletwo}%

\theoremstyle{thmstylethree}%

\begin{document}

\title[Radiation field analysis at MoEDAL using Timepix]{Analysis of mixed radiation fields at the MoEDAL~experiment based on real-time data from a Timepix detector network}


\author[1]{\fnm{Benedikt} \sur{Bergmann}\,\orcidlink{0000-0002-8076-5614}}
\author[1,2]{\fnm{Petr} \sur{Burian}\,\orcidlink{0000-0001-6907-7901}}
\author[1]{\fnm{Josef} \sur{Janeček}\,\orcidlink{0000-0002-5072-4141}}
\author[3,5]{\fnm{Claude} \sur{Leroy}\,\orcidlink{0000-0003-3105-7045}}
\author*[1]{\fnm{Petr} \sur{Mánek}\,\orcidlink{0000-0003-4306-0209}}\email{petr.manek@utef.cvut.cz}
\author[4]{\fnm{James} \sur{Pinfold}\,\orcidlink{0000-0002-2397-4196}}
\author[1]{\fnm{Stanislav} \sur{Pospíšil}\,\orcidlink{0000-0001-5424-9096}}
\author[4]{\fnm{Richard} \sur{Soluk}\,\orcidlink{0009-0003-8891-7648}}
\author[1]{\fnm{Michal} \sur{Suk}\,\orcidlink{0000-0002-0151-2388}}

\affil*[1]{\orgdiv{Institute of Experimental and Applied Physics}, \orgname{Czech Technical University in Prague}, \orgaddress{\street{Husova 240/5}, \city{Prague}, \postcode{110 00}, \country{Czech Republic}}}

\affil[2]{\orgdiv{Faculty of Electrical Engineering}, \orgname{University of West Bohemia}, \orgaddress{\street{Univerzitní 2732/8}, \city{Pilsen}, \postcode{301 00}, \country{Czech Republic}}}

\affil[3]{\orgdiv{Département de Physique}, \orgname{Université de Montréal}, \orgaddress{\street{1375 Avenue Thérèse-Lavoie-Roux}, \city{Montréal}, \postcode{QC H2V 0B3}, \country{Canada}}}

\affil[4]{\orgdiv{Department of Physics}, \orgname{University of Alberta}, \orgaddress{\street{4-181 Centre for Interdisciplinary Science}, \city{Edmonton}, \postcode{AB T6G 2E1}, \country{Canada}}}

\affil[5]{Deceased}


\abstract{The primary objective of this work is the determination of fluences
and characteristics of fast neutrons, other hadrons, and highly ionizing
particles~(HIPs) in the environment of the MoEDAL~experiment at the Large Hadron
Collider. These particles may constitute an experimental background for the passive
Nuclear Track Detectors~(NTDs) used by MoEDAL to search for tracks potentially
produced by magnetic monopoles, in particular by particles
indistinguishable in~NTD from monopoles. The study is based on data acquired by
the Timepix hybrid silicon pixel detector network, which represents the first
and only active detector system installed and operated as part of the MoEDAL
experiment from 2013 to 2018. The Timepix detector network enables real-time
measurements of mixed radiation fields, including the composition, spectral
properties, and directional characteristics of individual radiation components
across different regions of the MoEDAL experimental area. The paper presents
detailed results of the radiation field analysis with emphasis on neutrons and
HIPs, including their directional distributions. The first
results demonstrating the spatial tracking capabilities of the Timepix detectors
are also reported, illustrating the reconstruction of particle direction and
energy-loss profiles from individual detector frames.}




\maketitle

\section{Introduction}\label{sec1}

The MoEDAL experiment at the Large Hadron Collider~\cite{acharya2014physics} is
primarily designed to search for exotic highly ionizing particles~(HIPs), in particular
magnetic monopoles, using passive Nuclear Track Detectors~(NTDs). In the
experimental environment of MoEDAL, various background particles such as fast
neutrons, hadrons, and energetic ions are present and contribute to tracks
recorded by the NTDs. A quantitative determination of the fluence and
characteristics of these particles makes an important contribution to the
correct interpretation of NTD measurements.

To meet this requirement, a network of hybrid silicon pixel detectors based on
the Timepix technology~\cite{llopart2007timepix} was installed and progressively
commissioned in the MoEDAL experiment between~2013 and~2018. The deployment of
this system was motivated by the successful application of Medipix-type pixel
detectors in the
ATLAS~experiment~\cite{campbell2013analysis,heijne2022comparison}. The Timepix
detector network thus constituted the first active detector subsystem
implemented in MoEDAL.

Owing to the capability of Timepix detectors to register and track individual
interacting particles, the network enables detailed studies of mixed radiation
fields, with sensitivity to particle type, energy deposition, and flight
direction within the MoEDAL~experimental environment. Continuous real-time
operation of the system allowed monitoring of temporal and spatial variations of
the radiation field and enabled direct comparison with the exposure conditions
of the NTDs used in the experiment. In addition, this work presents first
results demonstrating the reconstruction of particle directions and energy-loss
profiles from individual particle tracks recorded in single detector frames.

\section{The Timepix detector in the MoEDAL environment}\label{sec2}

\subsection{The Timepix detector}\label{sec2sub1}

Timepix detectors consist of an active sensor layer
of~\qtyproduct[product-units=single]{1.4 x 1.4}{\cm\squared}. It is segmented
into a square matrix of 256~by~256~pixels with a pixel pitch of \qty{55}{\um}.
The sensor is bump bonded to the Timepix readout ASIC, which contains the
readout chain of each individual pixel to analyze the analog signal from the
charge sensitive preamplifier, discriminator and digital counter. The Timepix
detector relies entirely on a frame-based readout. A frame is defined as the
state of all pixel counters after a given integration time (frame-acquisition
time).

By reverse biasing the active sensor layer, a depleted volume is formed in which
ionizing radiation generates free charge carriers. These carriers drift through
the sensor under the electric field created by the applied potential difference
between the common backside and the pixel electrodes. Once they start their drift,
the charge carriers induce a signal at the pixel contacts, which is amplified,
shaped and compared to a globally adjustable threshold level
(\qty[parse-numbers=false]{\approx 5}{\keV}). Depending on the mode of
operation, the signal is interpreted
differently:%
\begin{itemize}
\item
In counting mode, the pixel counter is incremented once the amplified pulse
crosses the preset threshold level~(THL). The counter status thus represents the
number of interactions in a pixel.

\item
In Time-of-Arrival~(ToA) mode, the time is measured from the time the amplified
pulse crosses the THL~level until the end of the frame-acquisition time. The
maximum sampling frequency is~\qty{48}{\MHz}, giving \qty{23}{\ns} time
resolution. The counter status of the pixel thus relates to the particle arrival
time.

\item
In Time-over-Threshold~(ToT) mode, one measures the interval from the time when
the pulse crosses the THL~level on its upward slope until the time when it
crosses the THL on its downward slope. By a pixel-by-pixel calibration with
X-ray photons of known energy the ToT can be calibrated to
energy~\cite{jakubek2011precise}.
\end{itemize}

\subsection{Timepix network at MoEDAL}\label{sec2sub2}

\begin{figure}[h]
\centering
\includegraphics[width=0.9\textwidth,trim=-1.4bp 42.2bp 70.8bp 39.8bp,clip]{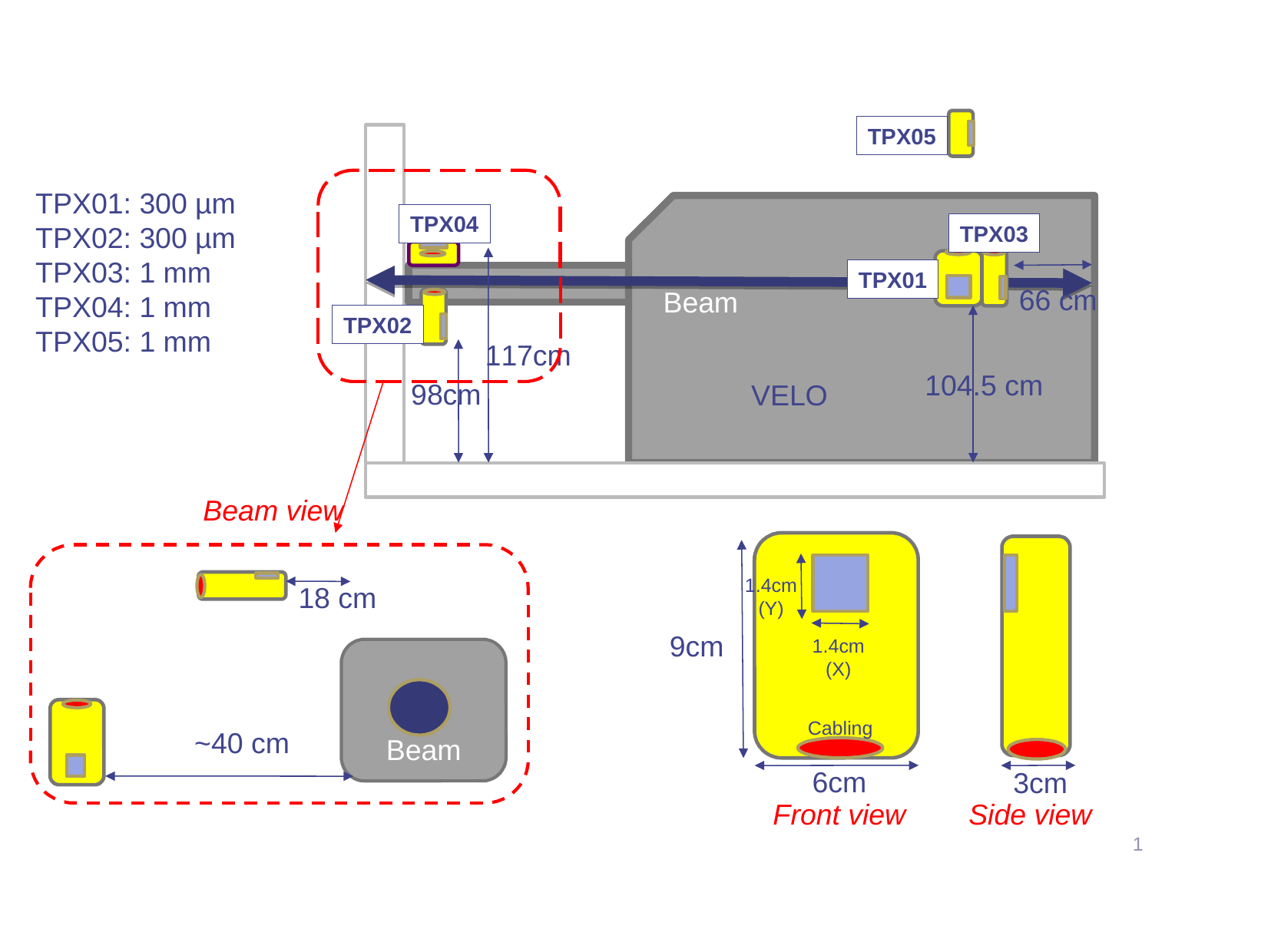}
\caption{Positions of the Timepix pixel detectors in MoEDAL.}\label{fig1}
\end{figure}

Five Timepix detectors are operated in MoEDAL. TPX01 and TPX02 are detecting
devices with \qty{300}{\um} thick silicon sensors, while TPX03, TPX04 and
TPX05 have a \qty{1}{\mm} thick silicon sensor layer. The positions are
shown in Fig.~\ref{fig1}. To increase neutron sensitivity, the surface of the
TPX05 detector is equipped with a set of neutron converters (\isotope[6]{Li}F, PE
(polyethylene), PE~+~\isotope{Al}); see further in Sect.~\ref{sec4}. It should be added that all
results presented in this work were obtained with Timepix detectors operated in
the Time-over-Threshold~(ToT) mode.

\begin{figure}[h]
\centering
\includegraphics[width=0.9\textwidth,trim=-5.8bp 56.5bp -18.9bp 25.6bp,clip]{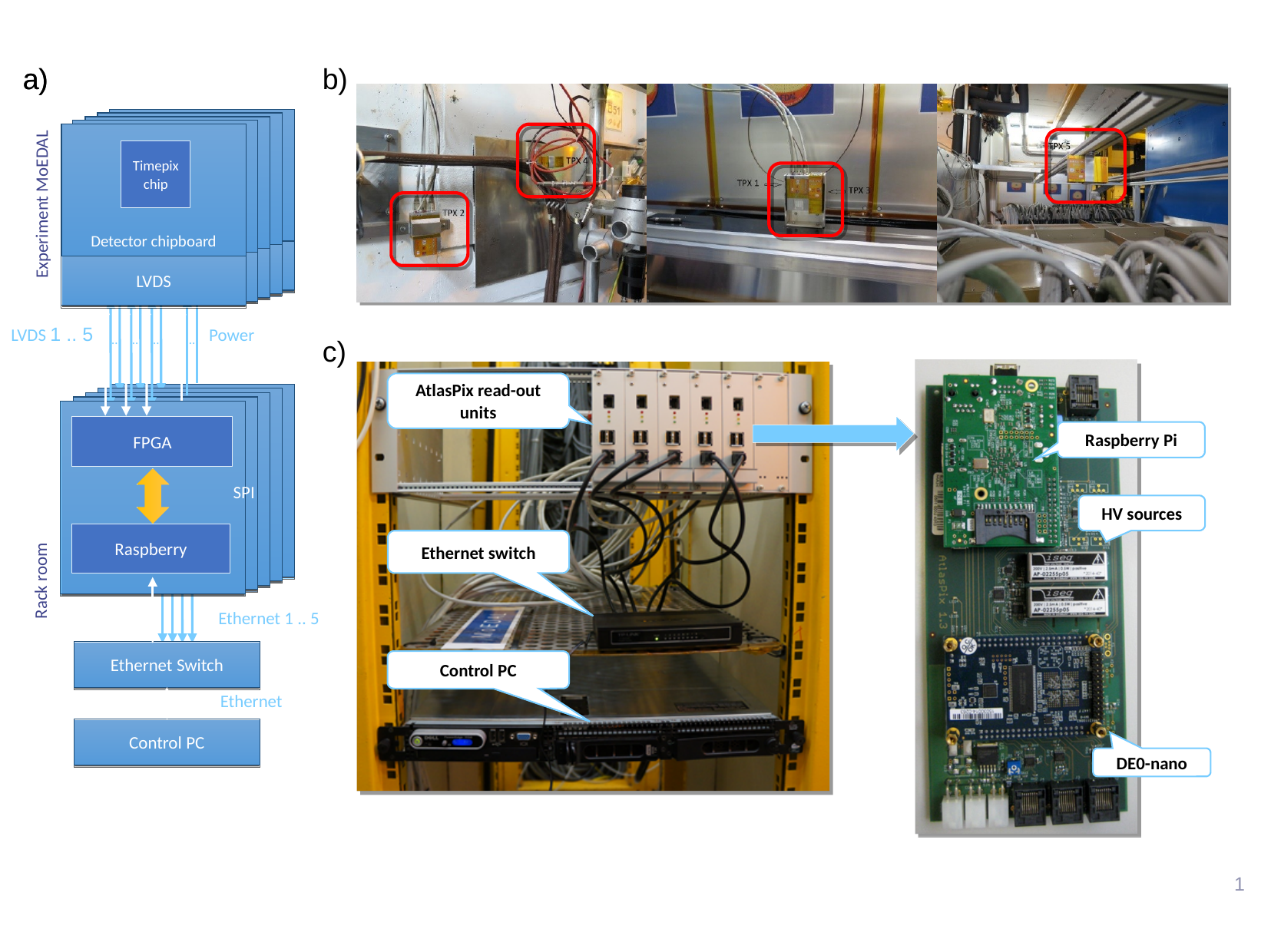}
\caption{a)~Scheme of the data flow; b)~Photographs of Timepix detector devices installed in the MoEDAL experiment; c)~Picture of the readout system and the board placed in the rack room.}\label{fig2}
\end{figure}

The data transfer scheme is shown in Fig.~\ref{fig2}. Each Timepix detector is
connected through four Ethernet cables to its corresponding readout device
(consisting of FPGA and Raspberry Pi mini-PC), which is located in the rack room
(in the D2~barrack), thus not being exposed to a high level of radiation. For
detector settings and data retrieval by the user, all 5 control units
communicate through an Ethernet switch with the control PC.

\section{Data analysis}\label{sec3}

\subsection{Cluster creation and basic pattern recognition}\label{sec3sub1}

With their high dynamic range and a detection threshold of~\qty{5}{\keV},
the Timepix detectors cover the measurement of a broad spectrum of particle
species and energies. It ranges from X-rays and low-energy electrons just above
the preset detection threshold to~HIPs.

\begin{figure}[h]
\centering
\includegraphics[width=0.7\textwidth]{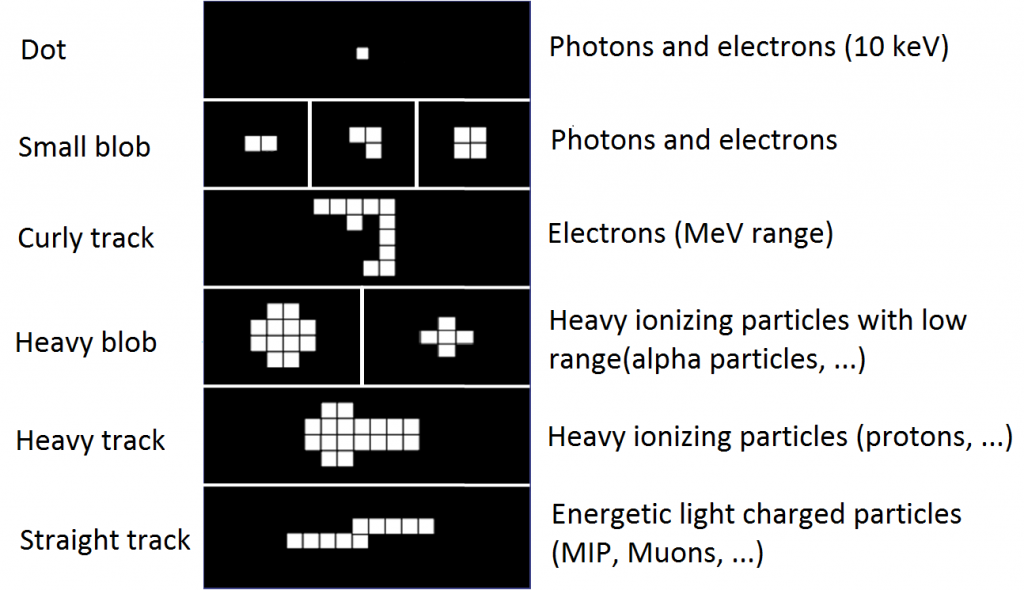}
\caption{Overview of defined basic cluster types and assignment of their
possible origins. Taken from~\cite{holy2008pattern}.}\label{fig3}
\end{figure}

The detector segmentation together with the charge transport properties of its
semiconductor sensitive layer allow a particle type differentiation by
evaluating their characteristic imprints (clusters/tracks) seen in the frames of
the pixel detector. A basic pattern recognition scheme was established
by~\cite{holy2008pattern} in~2008, which has been further developed to include
more particle species and energies. Ionizing radiation was classified into the
6 categories shown in Fig.~\ref{fig3}.

This categorization of the Timepix data can be seen as a starting point for
data analysis for the needs of the MoEDAL experiment. For a basic comparison to
the NTD~data, it is sufficient to separate~HIPs or high-energy transfer events
(seen as heavy tracks or heavy blobs) from low-energy transfer events.

For exotic particle searches, additional cluster parameters,
including energy loss~($\mathrm{d}E/\mathrm{d}x$), the number of outgoing
$\gamma$-rays, and the number of particle prongs, are considered. Therefore, the
Timepix detector responses to ionizing particles were studied in different ion
beams in the MeV and GeV energy ranges. These data are available as input for
advanced calibration of track pattern recognition of a large variety of
registered events.  An example of a particle with high velocity and stopping
power (\qty[per-mode=symbol]{75}{\GeV\per\clight} argon ion) measured at the
Super Proton Synchrotron at~CERN is shown in Fig.~\ref{fig4a}.
Figure~\ref{fig4b},~\ref{fig4c} shows star-like events produced in the sensor
layer of TPX03.

\begin{figure}[h]
\centering
\begin{minipage}[t]{0.3\textwidth}\centering
\includegraphics[width=\linewidth]{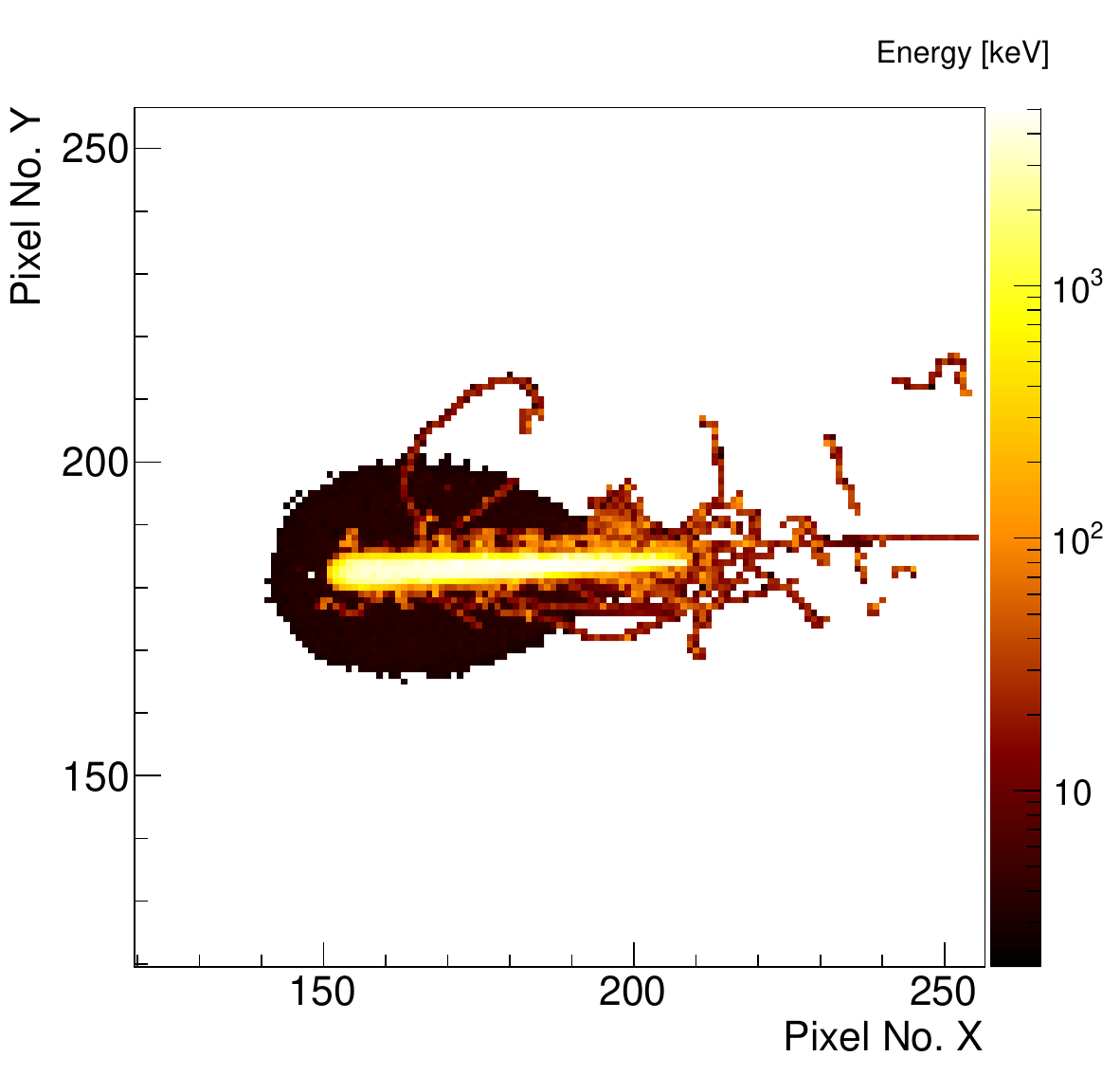}
\\\subfiglabel{fig4a}
\end{minipage}\hfill
\begin{minipage}[t]{0.3\textwidth}\centering
\includegraphics[width=\linewidth]{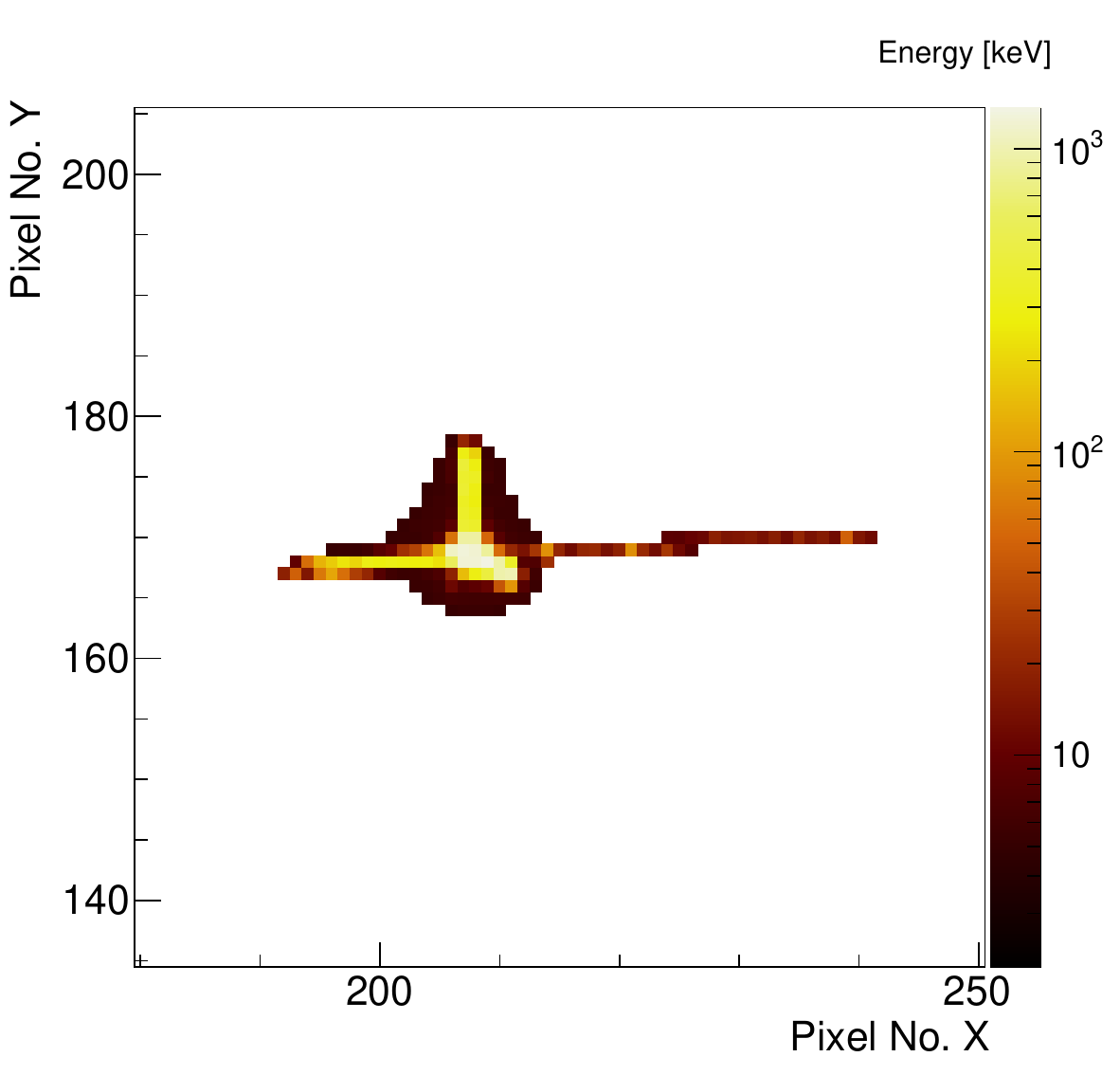}
\\\subfiglabel{fig4b}
\end{minipage}\hfill
\begin{minipage}[t]{0.3\textwidth}\centering
\includegraphics[width=\linewidth]{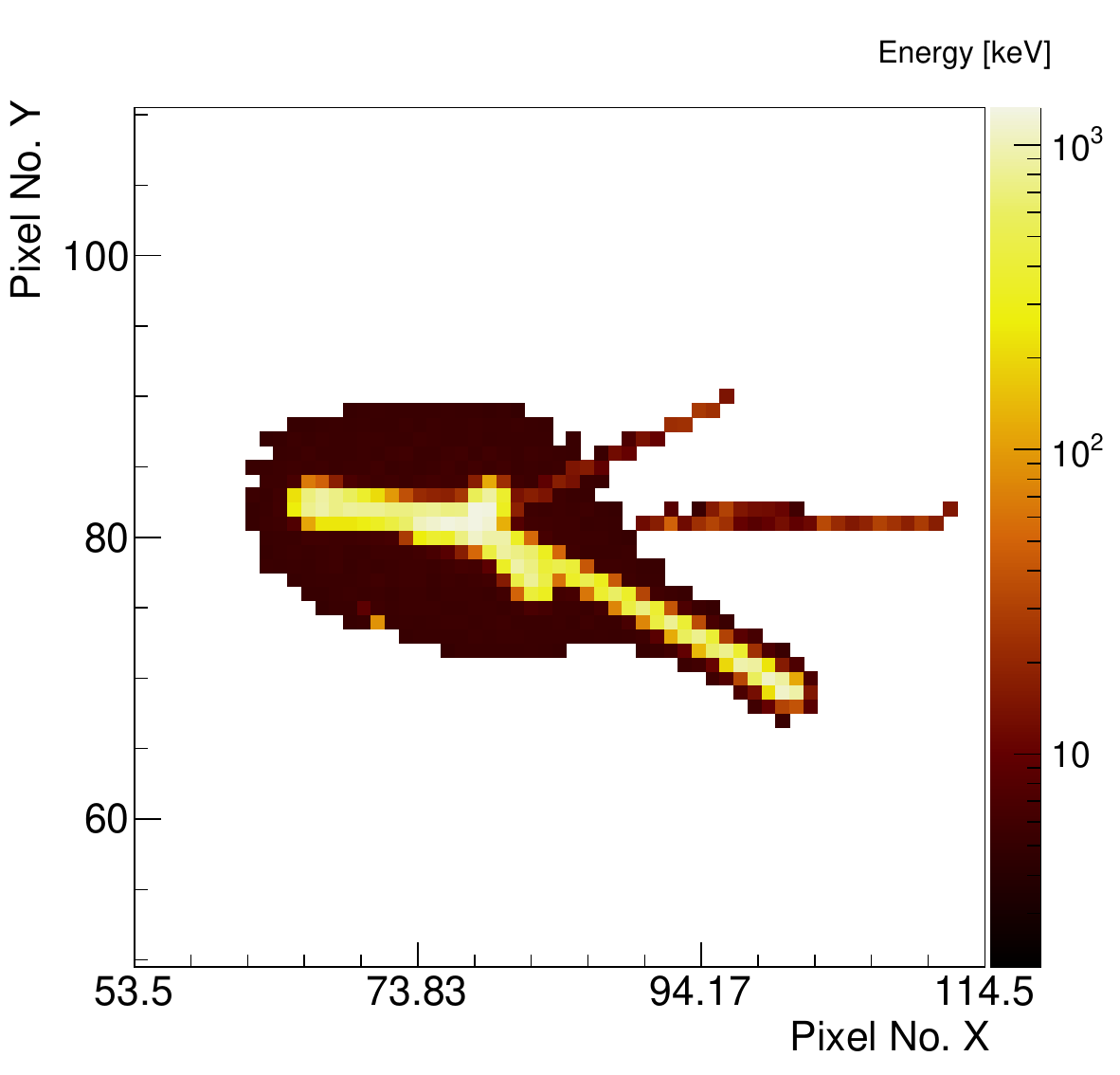}
\\\subfiglabel{fig4c}
\end{minipage}
\caption{Different events categorized as heavy tracks:
(a)~\qty[per-mode=symbol]{75}{\GeV\per\clight}~argon ion measured at the Super Proton Synchrotron~(SPS) at~CERN; (b)~and~(c) Star-like events observed with TPX03 in MoEDAL experiment. A subcategorization of star-like events and events with high $\delta$-ray multiplicity is currently in progress.}\label{fig4}
\end{figure}

\subsection{Composition of radiation fields at detector locations}\label{sec3sub2}

\begin{figure}[p]
\centering
\begin{minipage}[t]{0.4\textwidth}\centering
\includegraphics[width=\linewidth]{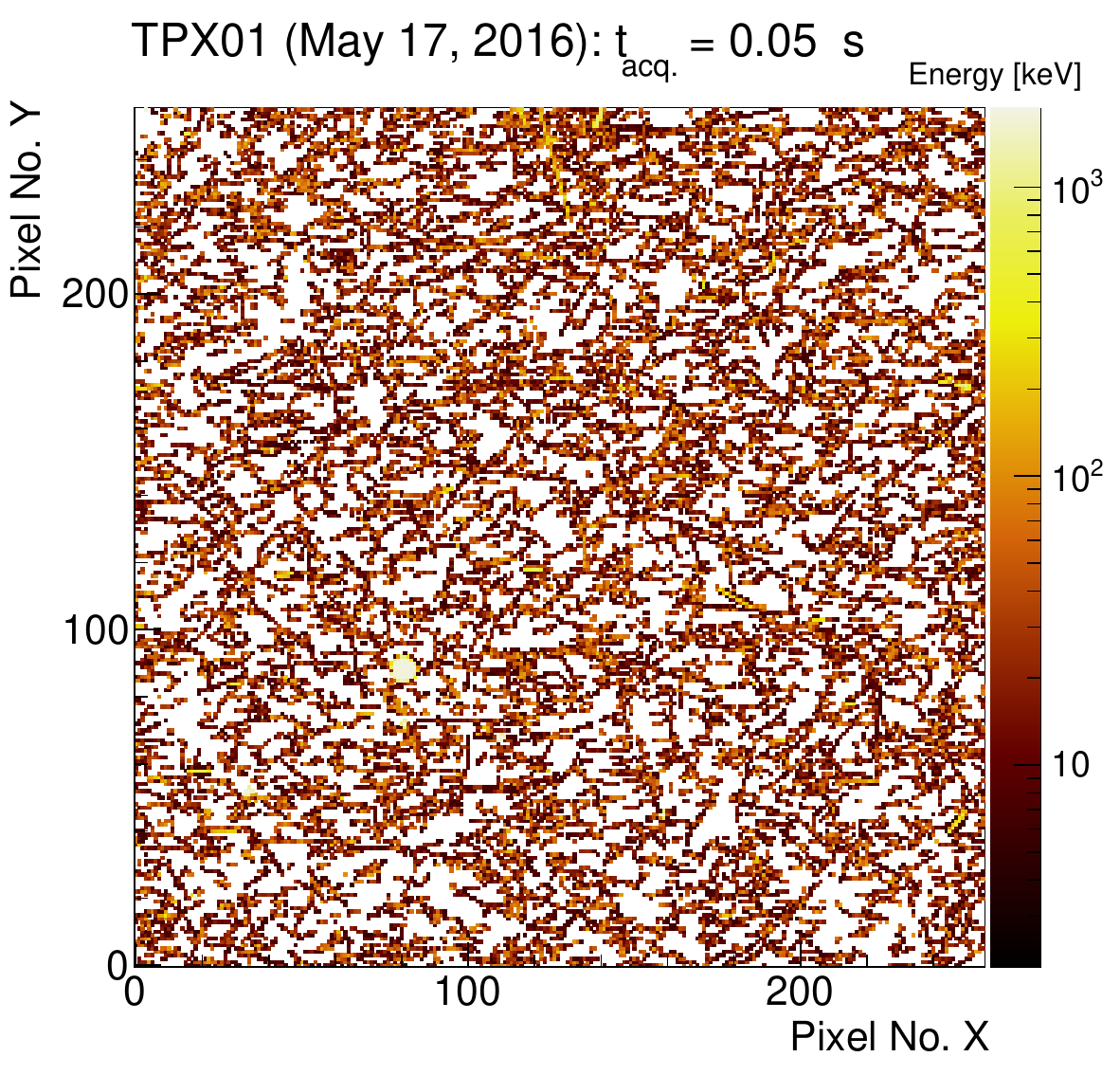}
\\\subfiglabel{fig5a}
\end{minipage}\qquad
\begin{minipage}[t]{0.4\textwidth}\centering
\includegraphics[width=\linewidth]{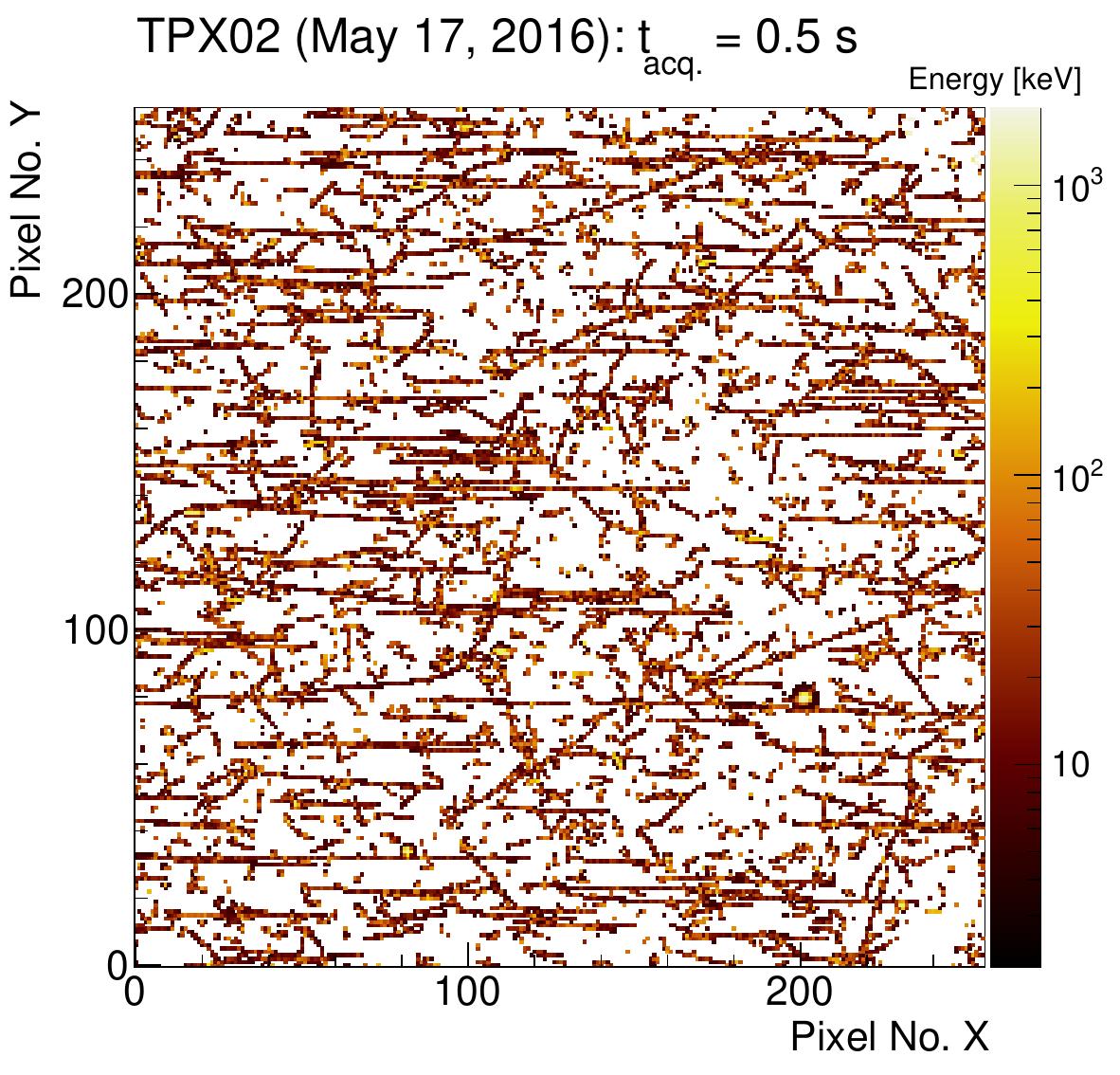}
\\\subfiglabel{fig5b}
\end{minipage}
\\
\begin{minipage}[t]{0.4\textwidth}\centering
\includegraphics[width=\linewidth]{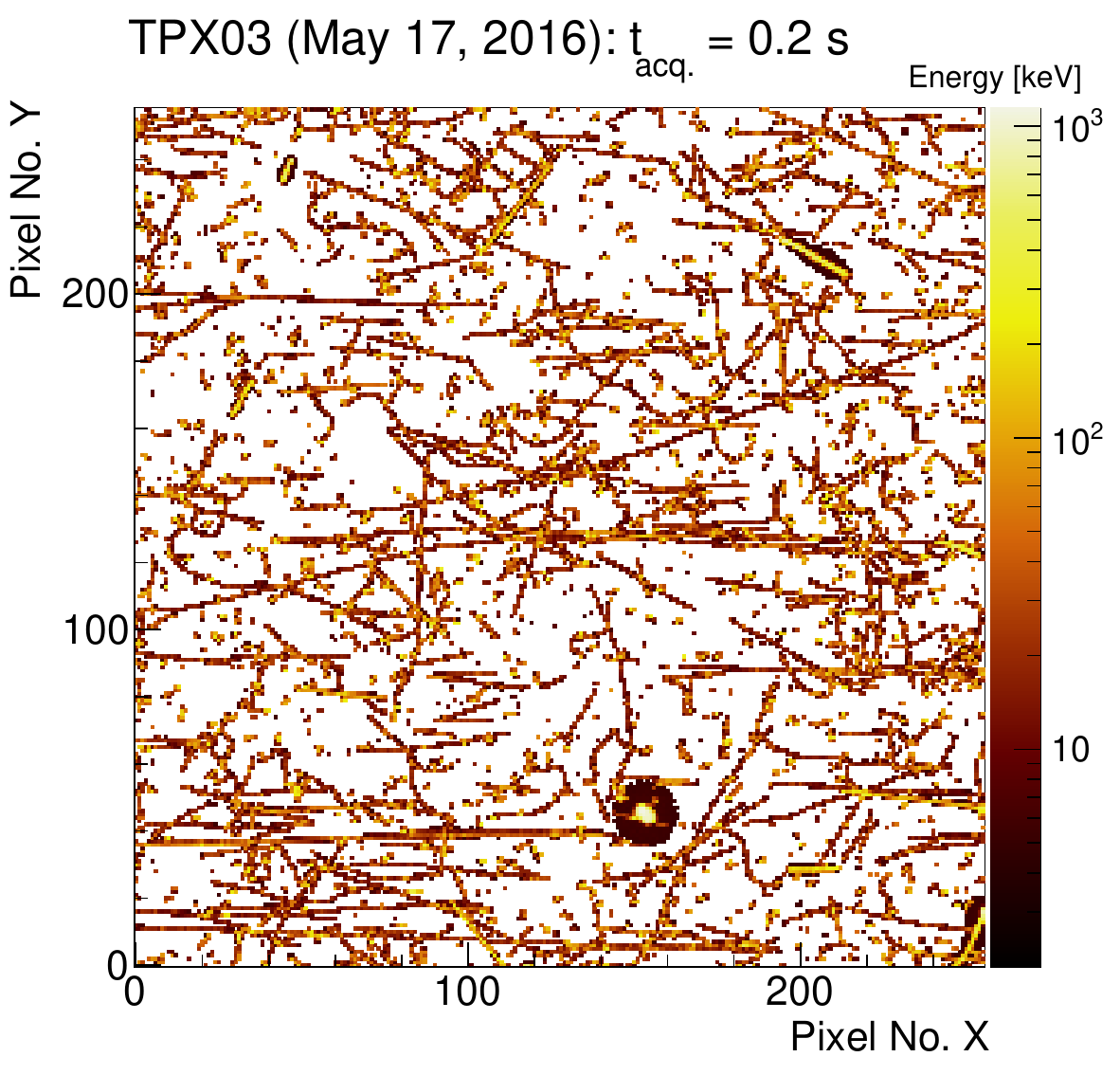}
\\\subfiglabel{fig5c}
\end{minipage}\qquad
\begin{minipage}[t]{0.4\textwidth}\centering
\includegraphics[width=\linewidth]{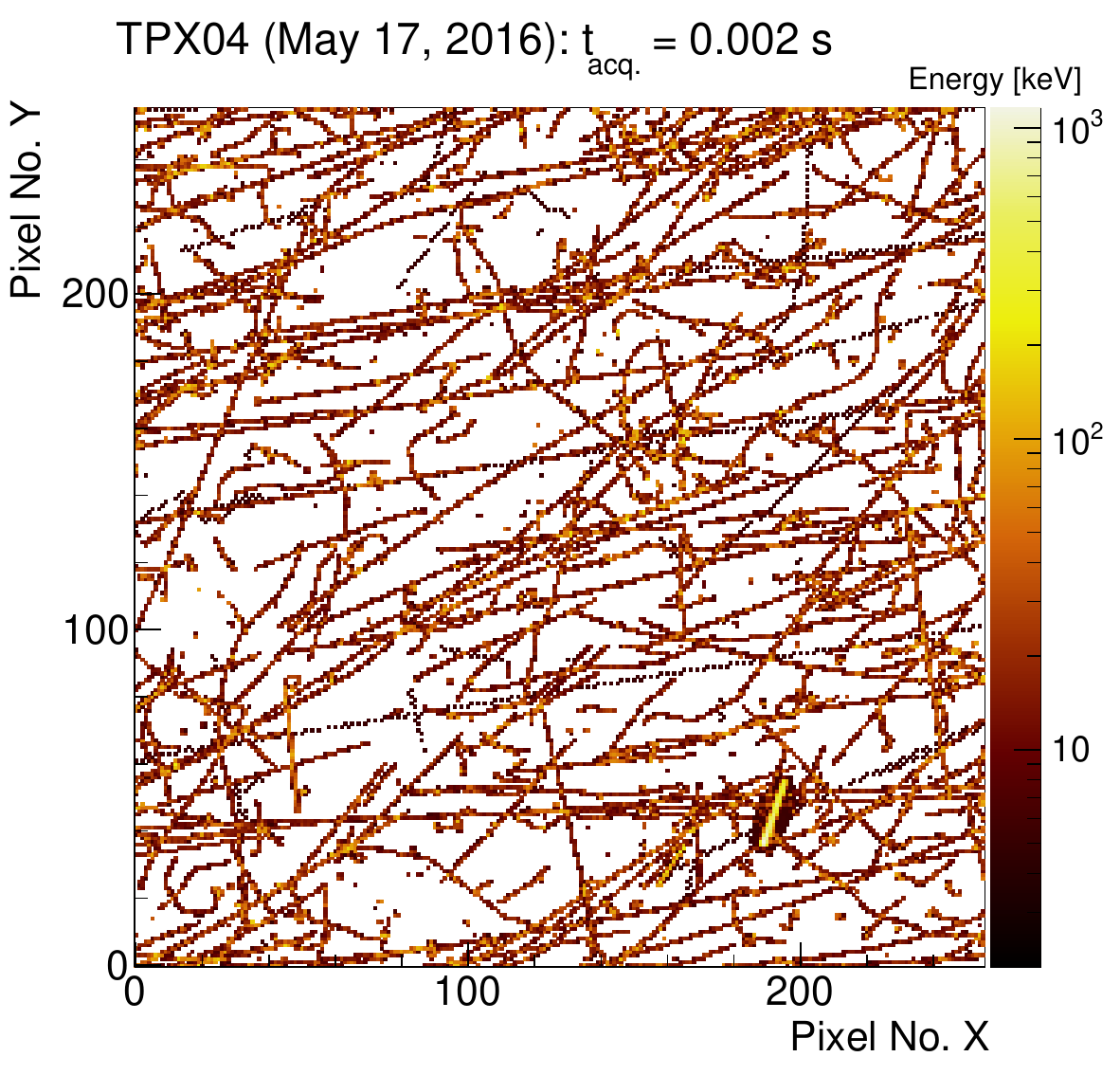}
\\\subfiglabel{fig5d}
\end{minipage}
\\
\begin{minipage}[t]{0.4\textwidth}\centering
\includegraphics[width=\linewidth]{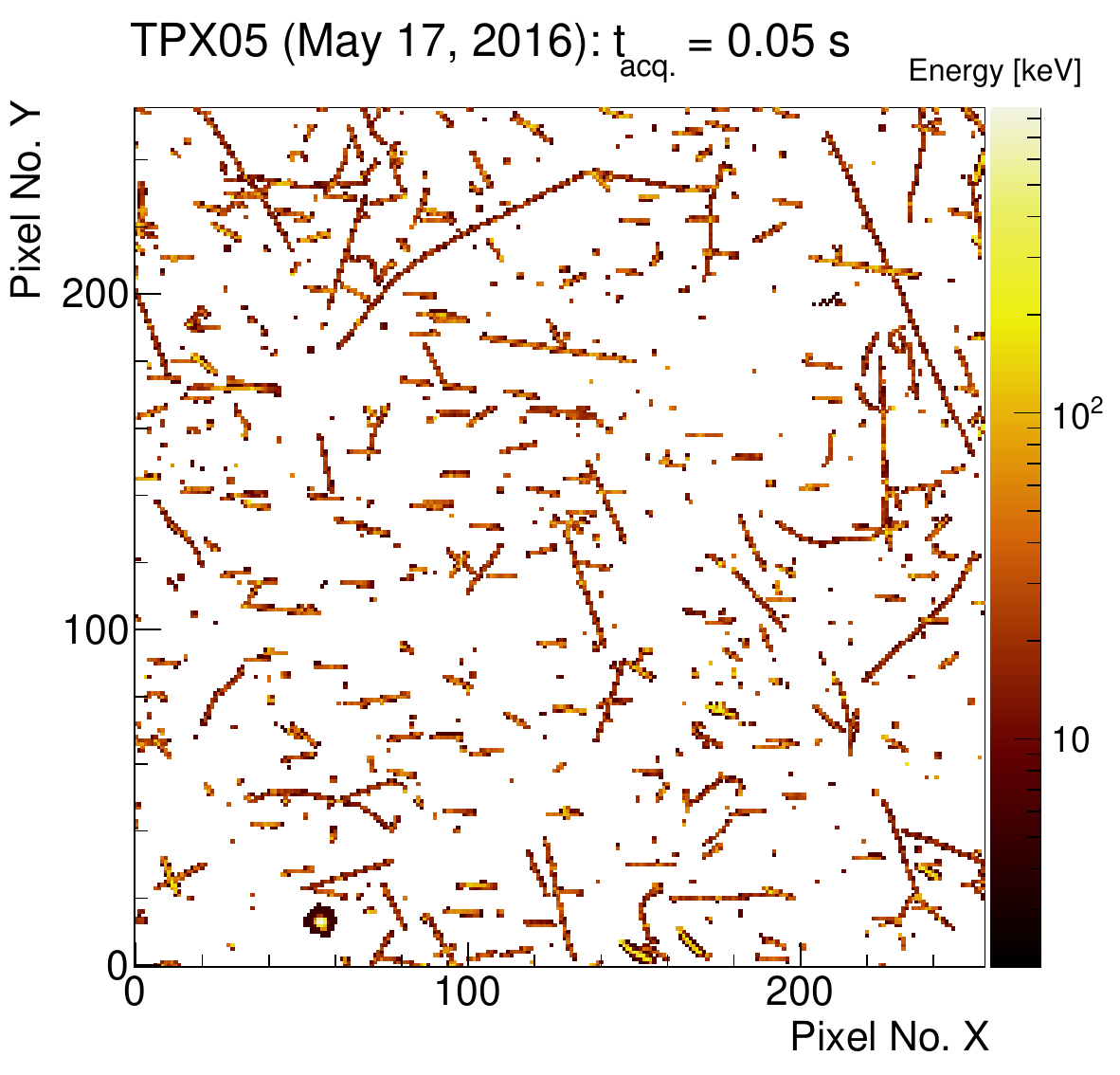}
\\\subfiglabel{fig5e}
\end{minipage}
\caption{Frames recorded by the Timepix detectors at various locations in the MoEDAL experimental environment illustrate the local radiation field composition. Distinct track patterns are observed depending on the detector position and its orientation with respect to the beam.}\label{fig5}
\end{figure}

The Timepix detector responses in the form of representative frames measured in
the MoEDAL environment are shown in Fig.~\ref{fig5}. The prevailing track length
and orientations depend on the detector thickness and orientation with respect
to the LHC beam pipe and interaction point of the LHCb.

\subsection{Identification of local mixed radiation field components using TPX03 data}\label{sec3sub3}

The data taken with TPX03 during proton-proton collisions
at~$\sqrt{s}=\qty{13}{\TeV}$ on May~17, 2016, are used as an example of how a mixed
radiation field component analysis can be approached. The fluence of high-energy
transfer events recorded by the detector per luminosity unit can be extracted
from these data. They represent a valuable input for analysis of the NTD~data
regarding the estimation of background signal caused by hadronic components
produced during beam-beam collisions.

A scatter plot of the energy deposition per track versus the cluster area (the
number of pixels within a cluster) is given in Fig.~\ref{fig6a}. It can be separated into
two regions:%
\begin{itemize}
\item Region 1, which contains particles with a low stopping power.
\item Region 2, which corresponds to particles with high-energy transfer to TPX03 silicon sensor.
\end{itemize}

A particle impact at a greater angle with respect to the sensor normal increases its
trajectory projection inside the pixel matrix, thus increasing the cluster area.
For particles with a low stopping power (region~1) such as minimally ionizing
particles (MIPs) and electrons, this increase relates to a small amount of
additional energy, so that these particles can be found on a branch with a small
slope in the scatter diagram (Fig.~\ref{fig6a}). A random subset of 300~tracks
seen in region~1 is given in Fig.~\ref{fig6b}. For~HIPs the
increase of particle trajectory is related to large additional energy. A further
analysis of track parameters can be performed, as shown by the example in
Fig.~\ref{fig6c}, where roundish shaped clusters were selected. Clusters with
significant deviation from a circle are shown in Fig.~\ref{fig6d}. The diversity
of individual tracks observed in Timepix detectors has motivated the development
of an analysis methodology for identification of neutron-induced events, which
is presented in the following section with references to published
literature~\cite{bergmann2016ionizing,bergmann2016atlas}, as well as a detailed
study of characteristic track signatures and their directional properties. These
features provide essential information on the origin of the detected particles,
and are discussed in the remainder of this work with reference to relevant
studies~\cite{Manek2018_CTU,bergmann20173d,bergmann2019characterization,bergmann2021sissa}.

\begin{figure}[h]
\centering
\begin{minipage}[t]{0.8\textwidth}\centering
\includegraphics[width=\linewidth]{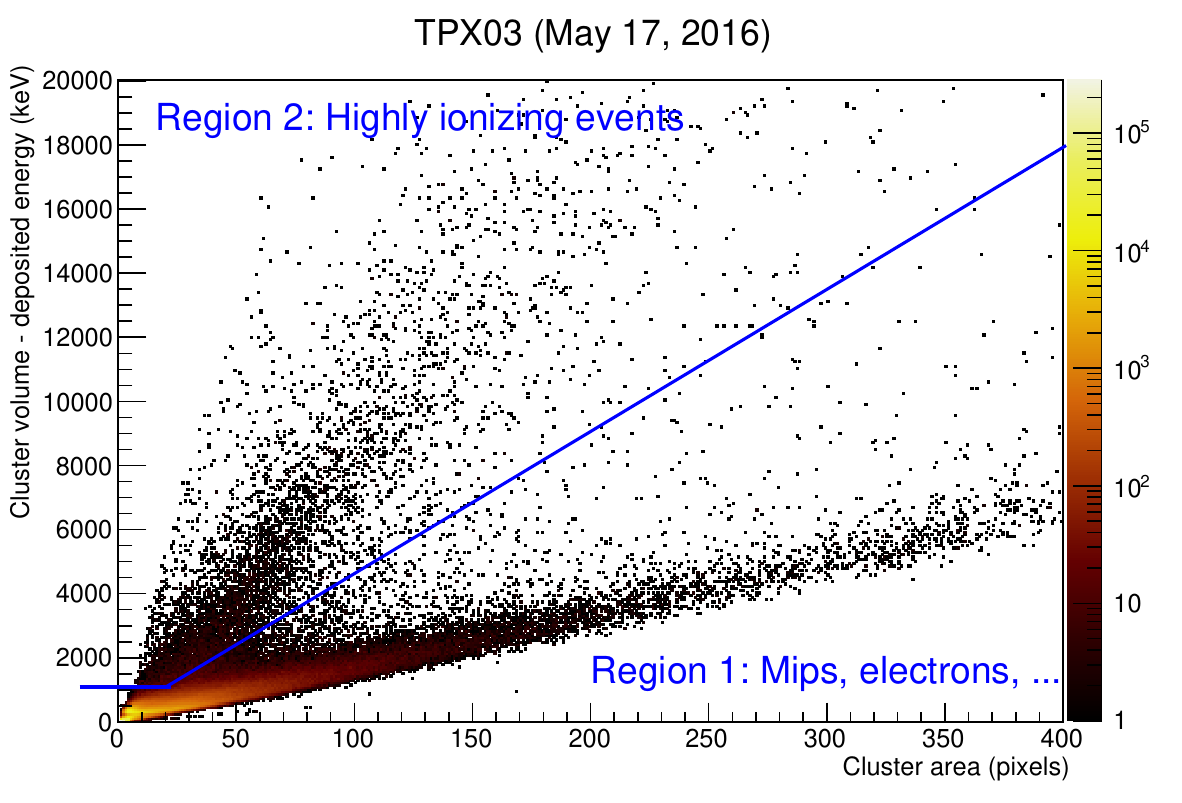}
\\\subfiglabel{fig6a}
\end{minipage}
\\
\begin{minipage}[t]{0.32\textwidth}\centering
\includegraphics[width=\linewidth]{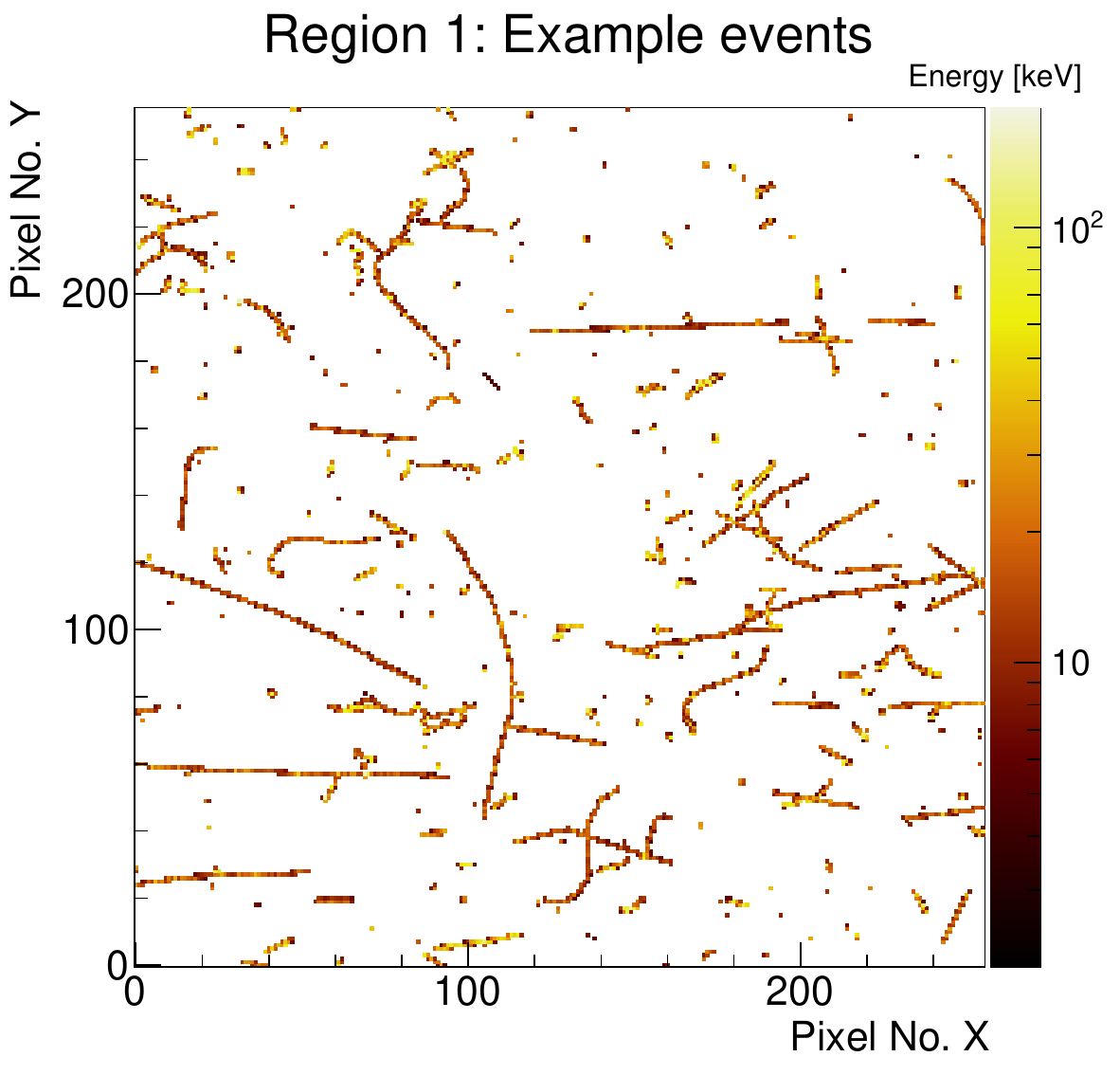}
\\\subfiglabel{fig6b}
\end{minipage}\hfill
\begin{minipage}[t]{0.32\textwidth}\centering
\includegraphics[width=\linewidth]{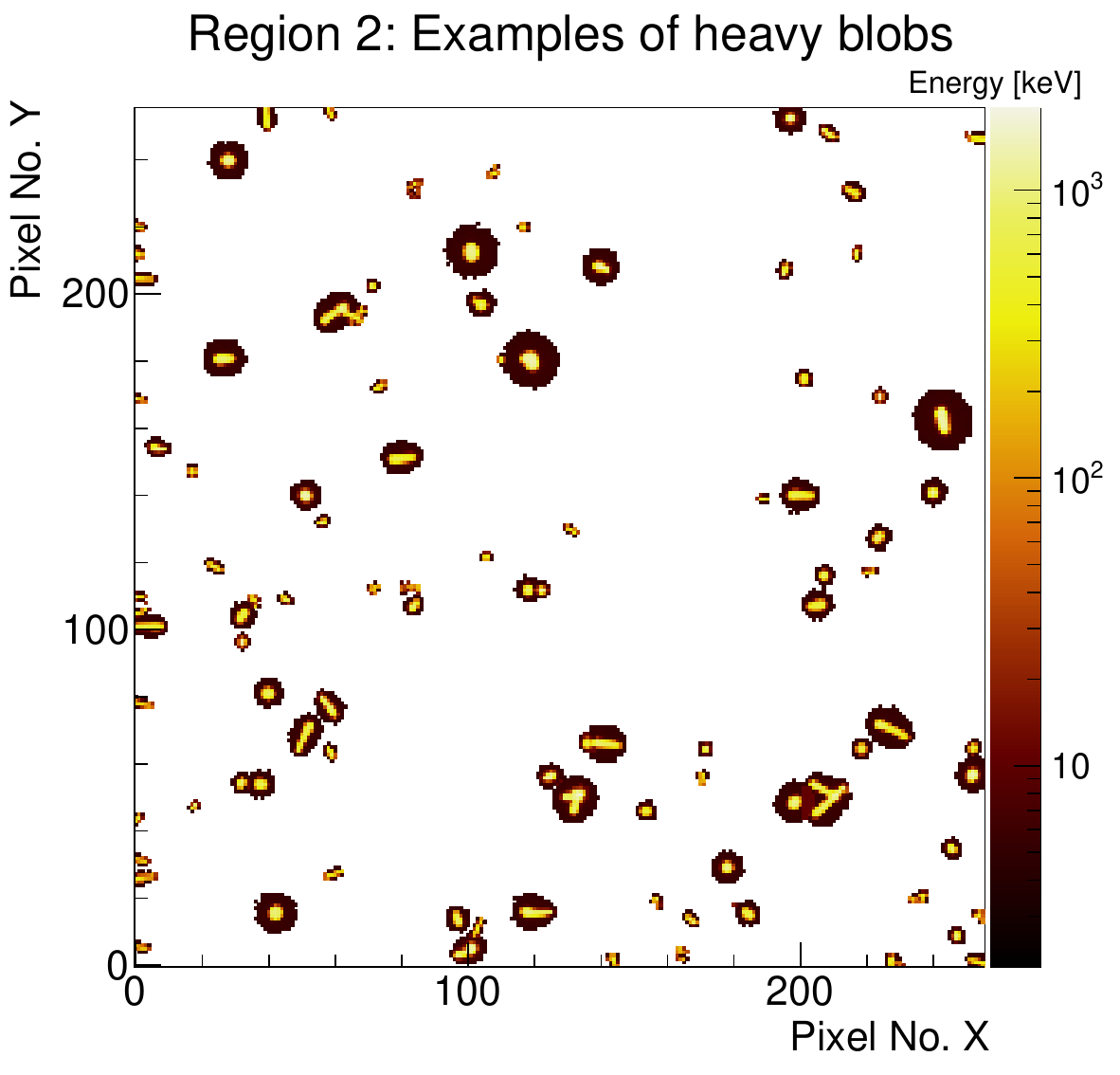}
\\\subfiglabel{fig6c}
\end{minipage}\hfill
\begin{minipage}[t]{0.32\textwidth}\centering
\includegraphics[width=\linewidth]{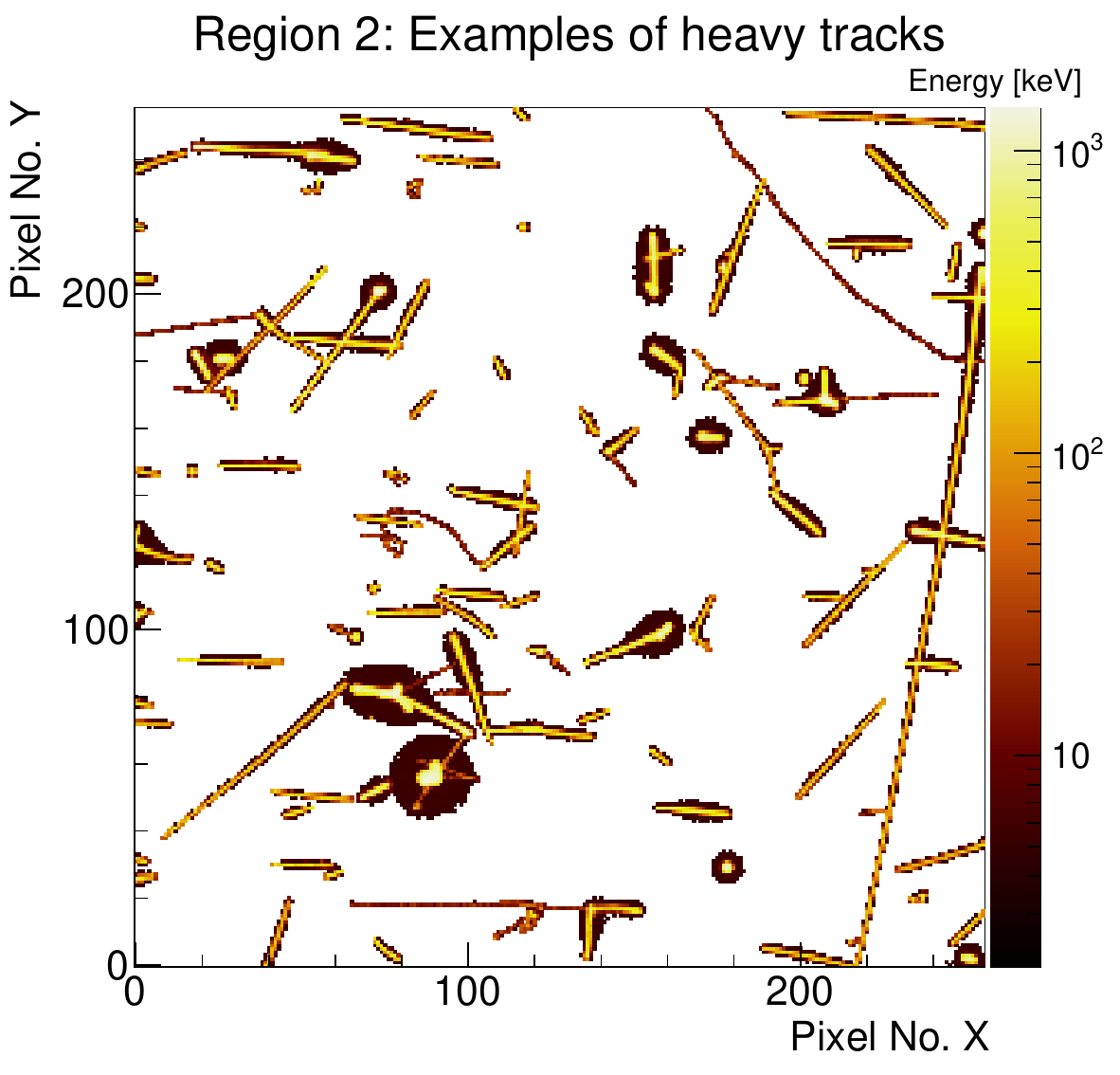}
\\\subfiglabel{fig6d}
\end{minipage}
\caption{(a)~Scatter plot of the deposited energy per cluster versus the cluster area obtained with TPX03. Two regions were defined (see text). (b)~Integral frame of 300~events randomly selected from region~1; (c)~Integral frame of 100~heavy blobs from region~2; (d)~Integral frame of 60~heavy tracks from region~2.}\label{fig6}
\end{figure}

\section{Neutron detection with TPX05}\label{sec4}

\subsection{Principles of neutron detection with Timepix detectors}\label{sec4sub1}

Neutrons do not produce charge carriers directly in the silicon sensor via
ionization processes. Thus, for their detection they have to be converted to
charged particles by means of specific converters. For fast neutrons, charge
carriers can also be generated directly in the silicon sensor through
neutron-induced nuclear recoils~\cite{bergmann2016ionizing}, $\isotope{Si}(n,\alpha)$ and
$\isotope{Si}(n,p)$ reactions (for neutron energies above
\qty[parse-numbers=false]{\approx 3}{\MeV}), and other neutron-induced
processes, including spallation (for neutron energies above \qty{15}{\MeV}).
However, the elastic and inelastic scattering of neutrons when the signal is
produced by recoil silicon nuclei cannot be efficiently used for neutron
detection due to their similarity to the signal of low-energy
photons~\cite{bergmann2016ionizing}.

Therefore, different regions of the pixel matrix are covered with converters for
thermal and increased fast neutron detection. The sensor matrix of~TPX05 is
separated into 4~different regions with almost equal area, each sensitive to
neutrons of different energies:%
\begin{itemize}
\item
\isotope[6]{Li}F~(\qty[quantity-product={}]{89}{\percent}~enriched
\isotope[6]{Li}, thickness: \qty[parse-numbers=false,per-mode=symbol]{\approx 1.3}{\mg\per\cm\squared}):
Thermal (with kinetic energy $T_n\approx \qty{25}{\meV}$) and epithermal neutrons
are detected through $\alpha$-particles and tritons from the
$\isotope[6]{Li}(n,\alpha)\isotope[3]{H}$ reaction (with thermal neutron cross-section
$\sigma=\qty[parse-units=false]{942}{\mathrm{barn}}$) and its well-known dependence on~$T_n$ mostly
obeying $1/(T_n)^{1/2}$ law, whereby the $\alpha$-particle and \isotope[3]{H} are
produced with the energy of~\qty{2.05}{\MeV} and~\qty{2.73}{\MeV},
respectively.
\item
Polyethylene~(PE, thickness: \qty[parse-numbers=false]{\approx 1.2}{\mm}): Fast neutrons
above~\qty{1}{\MeV} are detected through recoil protons from the PE-layer.
\item
PE~+~\isotope{Al} (thickness: \qty{1.2}{\mm} + \qty{80}{\um}): An
\qty{80}{\um}~thick aluminum foil is inserted between~PE and the sensor
layer to absorb the lower energy part of the neutron-recoil protons, and---in
this way---to create a region with a different energy threshold for neutron
detection. In this region, neutrons above \qty[parse-numbers=false]{\approx 4}{\MeV} are registered.
\item
Uncovered: The uncovered region is used to estimate the signal produced in the
silicon sensor itself and, consequently, to subtract it from responses below the
converters to obtain the net signals generated by neutrons in the converters, by
thermal neutrons in \isotope[6]{Li}F and fast neutrons in PE~regions.
\end{itemize}

The detection efficiency for thermal neutrons hitting the \isotope[6]{Li}F area was found
to be~\qty[parse-numbers=false,quantity-product={}]{\approx 0.5}{\percent}. For fast neutrons the detection efficiency in the~PE is in
the order of~\qty[parse-numbers=false,quantity-product={}]{\approx 0.1}{\percent} and strongly dependent on neutron energy. It shows a
maximum of~\qty[parse-numbers=false,quantity-product={}]{(0.32\pm0.02)}{\percent} at~\qty{16}{\MeV}. Comparing the responses
below the~PE and PE~+~\isotope{Al}~regions, the hardness of the neutron spectrum can be
assessed reliably in the energy region from~\qtyrange[range-units=single]{1}{20}{\MeV}~\cite{bergmann2016atlas}.

\subsection{Neutron fluence determination}\label{sec4sub2}

The determination of the neutron fluence is addressed with the example of the
TPX05~data from May~17, 2016. Figure~\ref{fig7a} shows an integrated frame of
neutron-like events (heavy blobs) for the full day
(around~\qty{27}{\per\nanobarn}). The regions covered by the different neutron
converter materials are indicated.

\begin{figure}[h]
\centering
\begin{minipage}[t]{0.41\textwidth}\centering
\includegraphics[width=\linewidth]{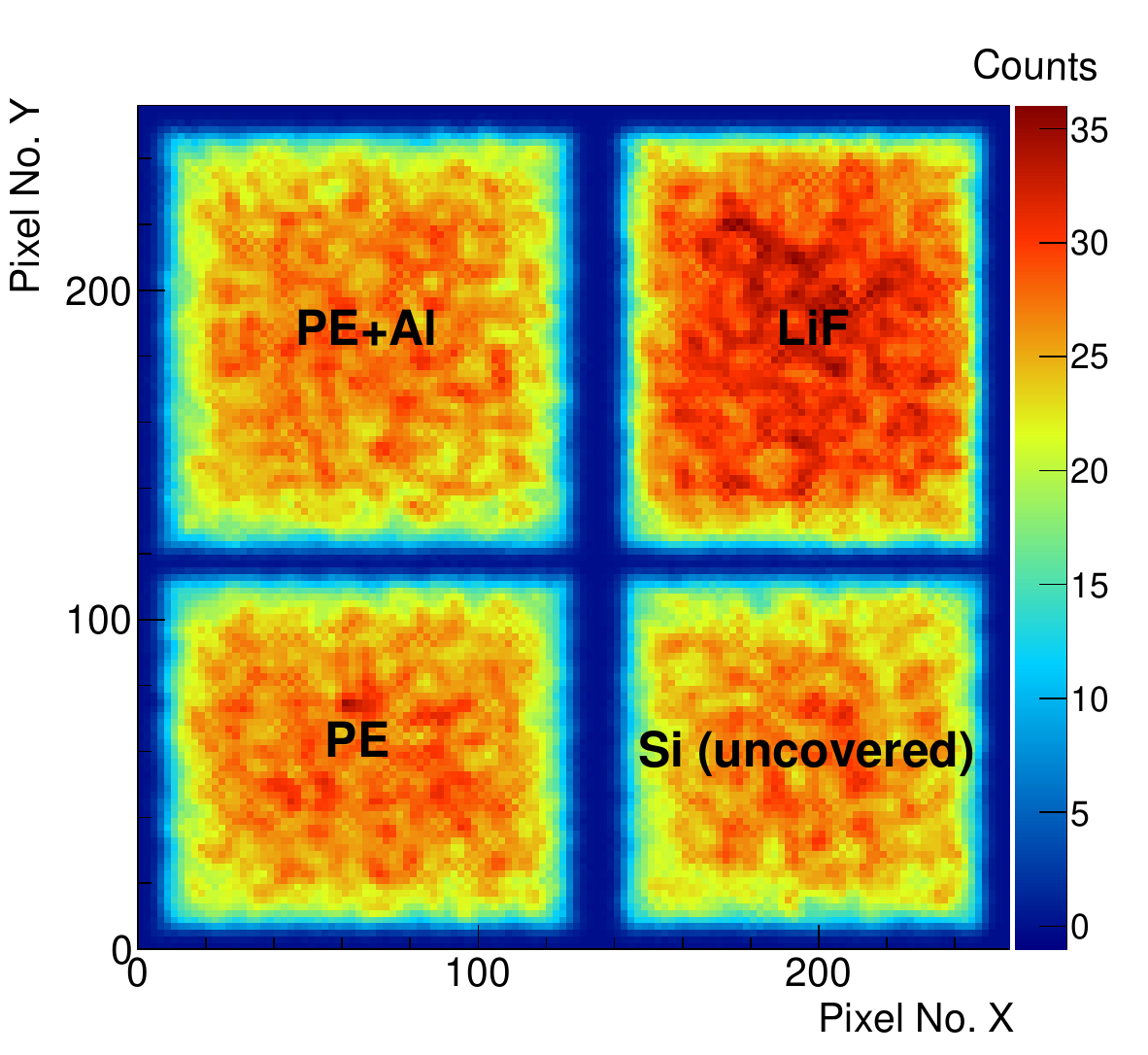}
\\\subfiglabel{fig7a}
\end{minipage}\hfill
\begin{minipage}[t]{0.56\textwidth}\centering
\includegraphics[width=\linewidth]{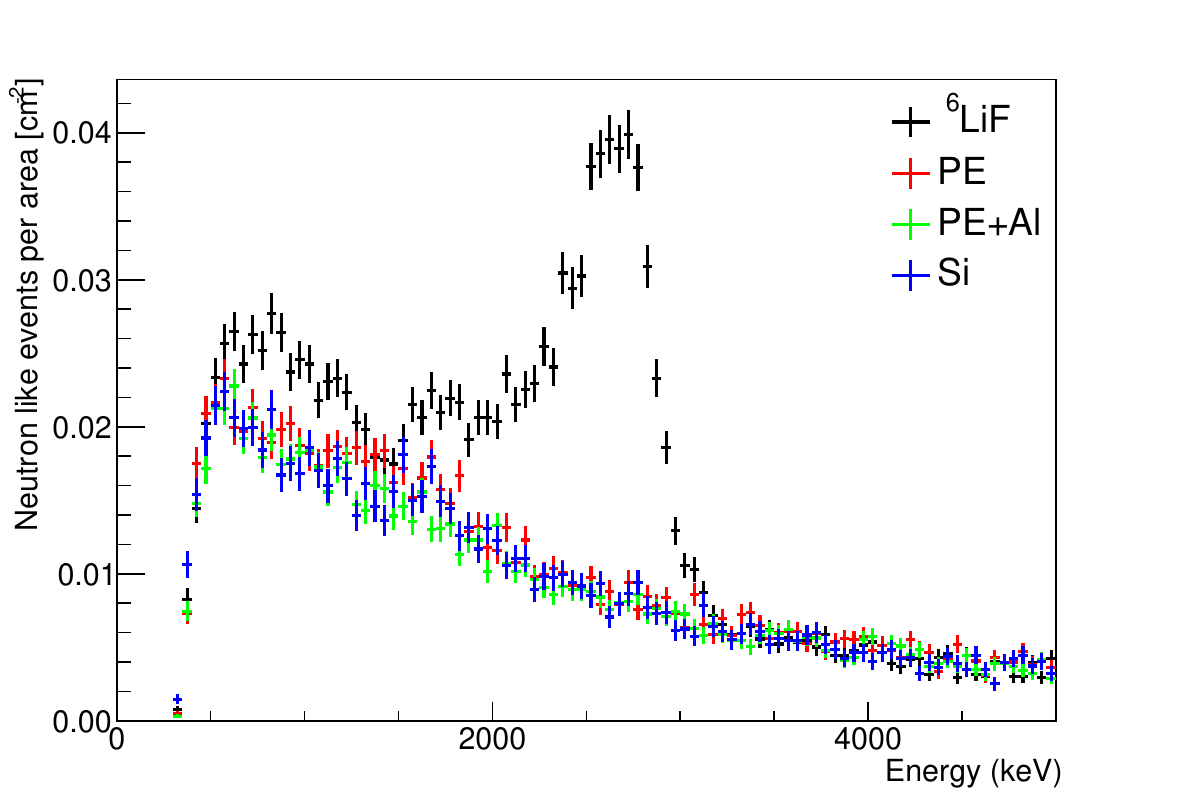}
\\\subfiglabel{fig7b}
\end{minipage}
\caption{(a)~Integral frame of neutron-like events measured with~TPX05; the regions covered with different converter materials are indicated. An increased response is clearly seen below the \isotope[6]{Li}F~layer, indicating the presence of thermal neutrons. For an unambiguous determination of the converter region, a masked area between the regions was defined and tracks with pixels in the masked region were excluded. (b)~Energy deposition of neutron-like events below the different converter regions. The energy spectrum of the $\isotope[6]{Li}(n,\alpha)\isotope[3]{H}$ reaction products is recognizable as two peaks (of alpha particles and tritons) above the signal of fast neutrons generated in the silicon sensor, thus permitting a reliable thermal neutron fluence measurement.}\label{fig7}
\end{figure}

An increased count rate below the \isotope[6]{Li}F~region can be seen, which indicates
the presence of a thermal neutron field. For a verification that the increase is
due to neutrons interacting in~\isotope[6]{Li}F, a comparison is made between the energy
deposition spectra below the different regions in Fig.~\ref{fig7b}. The~PE,
PE~+~\isotope{Al}, and the uncovered~(\isotope{Si}) region show a similar behavior, whereas in the
\isotope[6]{Li}F~region the peaks due to \isotope[3]{H}~(\qty{2.73}{\MeV}) and
$\alpha$-particle~(\qty{2.05}{\MeV}) energy depositions are obvious.

For a quantitative view, the average neutron fluxes are calculated according to%
\begin{align}
\Phi_{n,i} &= \frac{N_i/A_i - N_{\isotope{Si}}/A_{\isotope{Si}}}{\varepsilon_i\,N_\mathrm{frames}\, t_\mathrm{acq}}, \label{eq1}
\end{align}%
where $\varepsilon_i$ is the neutron detection efficiency, $N_\mathrm{frames}$
the number of measured frames with a frame-acquisition time of $t_\mathrm{acq}$,
$N_i$ the number of neutron-like events, and $A_i$ the area of the region~$i$
($i\in\{\mathrm{PE}, \mathrm{PE}\mathord{+}\isotope{Al}, \isotope[6]{Li}\mathrm{F}\}$). $N_{\isotope{Si}}$ and
$A_{\isotope{Si}}$ are the corresponding values for the uncovered region. The
uncertainty on the neutron flux is thus given by
\begin{align}
\Delta\Phi_{n,i} &= \sqrt{
\frac{N_i}{A_i^2\,\varepsilon_i^2\,N_\mathrm{frames}^2\,t_\mathrm{acq}^2}
+
\frac{N_{\isotope{Si}}}{A_{\isotope{Si}}^2\,\varepsilon_i^2\,N_\mathrm{frames}^2\,t_\mathrm{acq}^2}
+
\Phi_{n,i}^2\frac{\Delta\varepsilon_i^2}{\varepsilon_i^2}.
} \label{eq2}
\end{align}

The neutron detection efficiency is treated differently for fast and thermal neutrons. For thermal neutrons, a relative uncertainty of~\qty[quantity-product={}]{10}{\percent} is assumed, which corresponds to inhomogeneities within the batch of the produced
\isotope[6]{Li}F~layers. For fast neutrons, count rates show strong dependencies on the
kinetic energy and the impact angle of incident interactions~\cite{bergmann2016atlas} (see Fig.~10 therein). Consequently, this relationship has
to be taken into account in the analysis of anisotropic neutron fields. Given that
neither the neutron energy nor the impact angle is known at the location of~TPX05, it is
assumed that the fast component of the neutron field is dominated primarily by
albedo neutrons generated in the surrounding environment. Here, the detector
orientation admits selection of a representative range of impact angle
$\theta\approx\ang[parse-numbers=false]{(45\pm 25)}$; \textit{i.e.},
$\ang{20}-\ang{70}$. Furthermore, the kinetic energy of neutrons impacting the
PE~region is assumed to be at~\qty{16}{\MeV}, which
yields the largest possible efficiencies (\qty[quantity-product={}]{0.32}{\percent}~for the PE~region
and \qty[quantity-product={}]{0.25}{\percent}~for the PE+Al~region). The final step adds normalization by
average luminosity of~\qty{0.0406}{\per\nanobarn\per\s}, which corresponds to
LHCb Fills~4935 and~4937, resulting in fluences listed in Table~\ref{tab1}.

\begin{table}[h]
\caption{Overview of the values used for the determination of the total neutron
fluences $\Phi_n$ as measured with~TPX05 on May~17, 2016 (integrated
luminosity~\qty{27}{\per\nanobarn}). The converter detection
efficiencies~$\varepsilon_i$ were determined in~\cite{bergmann2016atlas};
$\theta$~is the impact angle of the neutrons; $T_n$~are the neutron kinetic
energy ranges to which the converters are sensitive. For scaling the neutron
fluxes the average luminosity of~\qty{0.0406}{\per\nanobarn\per\s} (part of LHCb
Fills~4935 and~4937) was used.}\label{tab1}%
\begin{tabular}{l r r r r}
\toprule
Region & \isotope[6]{LiF} & PE & PE+Al & Si \\
\midrule
$N_{i}$ & 88,504	& 73,213 & 86,173 & 62,858 \\
$A_{i}$ [cm$^{-2}$] & 0.440 & 0.423 & 0.499 & 0.373 \\
$\varepsilon_{i}$ [\%]~\cite{bergmann2016atlas} & $(0.50 \pm 0.05)$ &  $0.32$ &  $0.25$ & - \\
$\theta$ coverage & full &  $\ang[parse-numbers=false]{(45\pm 25)}$ &
$\ang[parse-numbers=false]{(45\pm 25)}$ & - \\
\midrule
T$_{\rm n}$ & $\approx 25$\,meV & $>1$\,MeV &  $>4$\,MeV & - \\
$\bm{\Phi_{n}}$
$\left[\frac{\mathrm{\bf{cm}}^{\bm{-2}}}{\mathrm{\bf{nb}}^{\bm{-1}}}\right]$ & $\bm{\left(12.1 \pm 1.3\right) \times 10^{3}}$ & $\bm{\left(3.7^{+3.9}_{-0.9}\right) \times 10^{3}}$ & $\bm{\left(3.4^{+3.7}_{-0.8}\right) \times 10^{3}}$ & -  \\
\bottomrule
\end{tabular}
\end{table}





\section{3D trajectory reconstruction}\label{sec5}

This section presents a demonstration of spatial tracking capabilities of the
Timepix detector network at MoEDAL. The principal aim is to show that
three-dimensional direction and energy-loss profile of individual particles
traversing sensors can be recovered from Timepix frames. Verification is carried
out by comparing the resulting per-detector distributions with the well-known
geometry of the LHCb cavern reported in Sect.~\ref{sec2sub2}.

\subsection{Methodology}\label{sec5sub1}

Reconstruction of three-dimensional particle trajectories from Timepix~frames
proceeds in three stages: cluster construction, overlap separation, and
trajectory fitting. Following that, energy loss is sampled along reconstructed
trajectories to estimate stopping power. The presented workflow is built around
the algorithms developed in~\cite{Manek2018_CTU}, which provides detailed
description of methods and values of parameters used by this analysis.

\begin{figure}[h]
\centering
\includegraphics[width=0.9\textwidth,trim=0bp 169.5bp 0bp 0bp,clip]{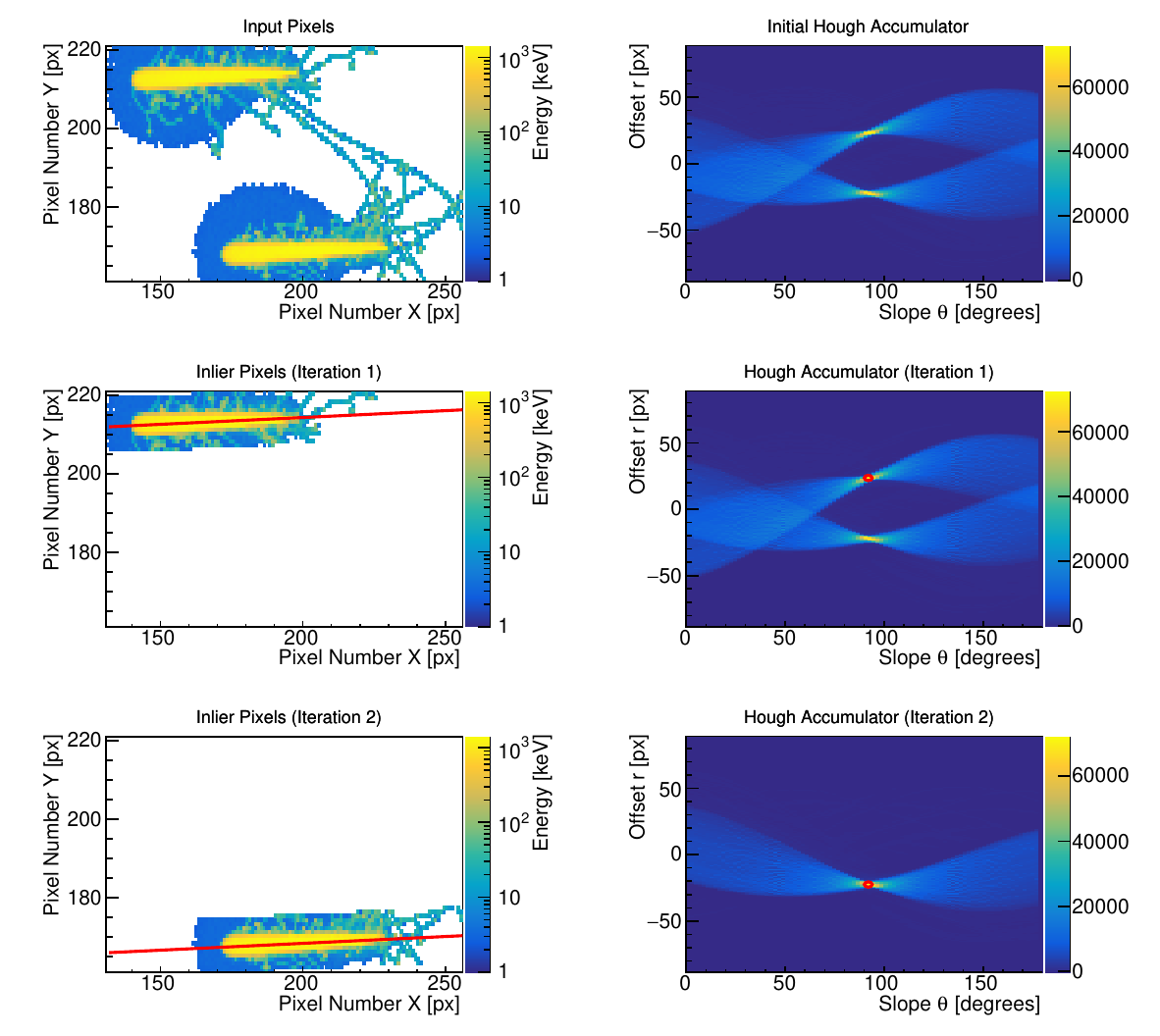}
\caption{Example of cluster overlap resolution by means of Hough
transformation for lines. A single cluster composed of 2~coincident overlapping argon tracks
is iteratively segmented into 2~groups. At each iteration, the cluster~(left) is
projected into the Hough domain~(right), where accumulator cells represent
2D~lines. Following that, the global Hough maximum is identified~(indicated by
a red circle), its supporting hits are subtracted and treated separately from that
point onward. Reproduced from~\cite{Manek2018_CTU}.}\label{fig8}
\end{figure}

Cluster construction is mostly consistent with the description given in
Sect.~\ref{sec3sub1}. Since this method alone cannot detect random overlaps, in
which two or more coincident tracks share pixels, each cluster is further
analyzed by Hough transformation for lines~\cite{HoughLineTransform},
parameterized in Hesse normal form. Lines are extracted iteratively by greedy
detection of the global accumulator maximum, identification of its inlier
pixels, and subtraction of their votes (Fig.~\ref{fig8}). This procedure repeats
until the maximum drops below a set threshold or after a sufficient number of
iterations is performed. Hits that are found to support distinct maxima within a
single cluster are analyzed separately.

Once overlaps are resolved, the final step of reconstruction focuses on
trajectory fitting. A particle trajectory within the silicon sensor is
approximated as rectilinear, which is adequate for the momentum of interacting
particles that dominate the analyzed data sample. Each trajectory is
parameterized by its intersection $\vec{x}_\text{far}$ with the back surface of
the sensor, an azimuth $\varphi$, and an incidence angle $\theta$ with respect
to the sensor normal~(sketched in Fig.~\ref{fig9a}). This parameterization has
the two well-known ambiguities arising from swaps of entry and exit points,
which are not expected to significantly affect the angular distributions
reported in this work.

The parameters $(\vec{x}_\text{far},\varphi,\theta)$ are fitted by a
non-deterministic RANSAC~algorithm: hit pixel pairs are drawn from the cluster
at random, each such pair determining a candidate trajectory. The trajectory
maximizing a likelihood metric over the cluster pixels is retained until a
better candidate is found, or the algorithm terminates. Each accepted candidate
is further refined by a local optimizer. For the energy-loss analysis presented
in Sect.~\ref{sec5sub4}, each fitted trajectory is additionally sampled along its
length in 128~equal segments using a neighborhood-suppressed estimator, which
integrates pixel energies in a transverse band around the trajectory while
suppressing contributions from low-energy pixels.

\subsection{Experimental data and processing}\label{sec5sub2}

The analysis presented in this work uses a series of LHCb runs from a \qty{3}{\hour}
time period on the night of July~29--30, 2017,\footnote{Taken during LHCb
Runs~195950, 195952, 195953, 195956 and 195958 (LHC Fill~6024).} during
which all 5~detectors were operational. Frame durations in the analyzed runs
were configured to \qtylist[list-units=single]{1;10;5;0.02;5}{\ms} for
detectors TPX01--TPX05, respectively.

Frames were filtered for sufficient data quality: only frames with occupancy
lower than~\qty[quantity-product={}]{15}{\percent} were accepted, and clusters that
were morphologically classified as straight or heavy tracks (definition in
Fig.~\ref{fig3}) were selected in order to maximize angular resolution. The
resulting collection of frames (example shown in Fig.~\ref{fig9b}) was
split into clusters and fitted by the trajectory reconstruction methods
described in Sect.~\ref{sec5sub1}. This process yielded several hundred thousand
trajectories per detector (exact quantities listed in Table~\ref{tab2}), which represent the data sample presented in the following sections.

\begin{figure}[h]
\centering
\begin{minipage}[t]{0.55\textwidth}\centering
\includegraphics[width=\linewidth]{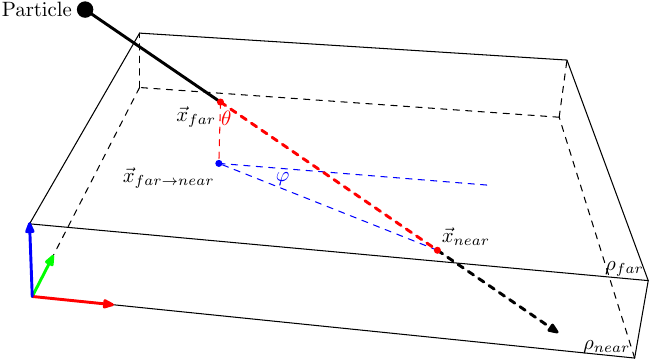}
\\\subfiglabel{fig9a}
\end{minipage}\hfill
\begin{minipage}[t]{0.38\textwidth}\centering
\includegraphics[width=\linewidth,trim=190bp 0bp 188bp 0bp,clip]{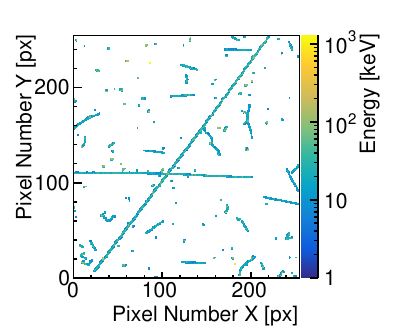}
\\\subfiglabel{fig9b}
\end{minipage}
\caption{(a)~Sketch of the trajectory parameterization $(\vec{x}_\text{far},\varphi,\theta)$.
The thick black arrow marks a rectilinear particle trajectory through the sensor
layer; the incidence angle~$\theta$~(red) is closed by the trajectory and the
sensor normal, the azimuth~$\varphi$~(blue) by the trajectory's projection into
the sensor plane and the $X$-axis. (b)~Example of filtered frames reported
by~TPX03, containing clusters that represent trajectories of minimally ionizing
particles. Such trajectories are later fitted by Hough transformation and
RANSAC. Reproduced from~\cite{Manek2018_CTU}.}\label{fig9}
\end{figure}

\subsection{Angular parameter analysis}\label{sec5sub3}

The strongest source of ionizing radiation in the experimental region is the
LHCb~interaction point itself, so the bulk of reconstructed trajectories is expected to point back
toward it. To test this, fitted trajectories are aggregated and binned per detector in
the two angular parameters of the trajectory model: the azimuth
$\varphi\in[0,\pi]$ and the incidence angle $\theta\in[0,\pi/2]$. The
resulting two-dimensional histograms are plotted in Fig.~\ref{fig10a},~\ref{fig10c}.
Owing to the cyclical nature of $\varphi$, the left and right borders of each
panel are continuous (\textit{i.e.}, as if the histogram were drawn on a
cylindrical surface).

The expected dominant trajectory direction in each detector follows directly
from its position and orientation with respect to the beam pipe, as noted in
Sect.~\ref{sec2sub2} and shown in Fig.~\ref{fig1}: a frontal detector is expected
to record orthogonal incidence ($\theta\approx0$); a detector facing the beam
from the side at the same height records trajectories with $\varphi\approx0$ and
$\theta$ controlled by the beam-to-detector geometry; and a detector parallel to
the beam records near-grazing incidence ($\theta\approx\pi/2$) with $\varphi$
set by the transverse offset. The expected and observed dominant peaks are
summarized in Table~\ref{tab3}: in all five detectors, the observed peaks are
found to be consistent with the geometric expectation. Since detectors were
installed close to NTDs, their results are useful for interpretation of
NTD~observations.

\begin{table}[h]
\centering
\begin{minipage}[t]{0.28\textwidth}
\centering
\caption{Number of fitted particle trajectories per Timepix~detector.}\label{tab2}%
\setlength{\tabcolsep}{3pt}%
\begin{tabular}{@{}lr@{}}
\toprule
Detector & Trajectory count\\
\midrule
TPX01 & 932,416\\ 
TPX02 & 587,421\\ 
TPX03 & 447,720\\ 
TPX04 & 869,311\\ 
TPX05 & 630,322\\ 
\bottomrule
\end{tabular}
\end{minipage}\hfill
\begin{minipage}[t]{0.70\textwidth}
\centering
\caption{Expected dominant trajectory directions per detector (from the
geometric position with respect to the LHC beam pipe, see Fig.~\ref{fig1}) and the
dominant peaks observed in Fig.~\ref{fig10a},~\ref{fig10c}.}\label{tab3}%
\setlength{\tabcolsep}{3pt}%
\begin{tabular}{@{}lll@{}}
\toprule
Detector & Expected $(\varphi,\theta)$ & Observed peak $(\varphi,\theta)$\\
\midrule
TPX01 & no preferred $\varphi$, $\theta\approx0$ (frontal)         & scattered $\varphi$, $\theta\in\{0,\pi/6\}$ \\
TPX02 & $\varphi=0$, $\theta\in[\pi/4,\pi/2]$                  & $\varphi=0$, $\theta\in[\pi/4,\pi/2]$ \\
TPX03 & $\varphi=0$, $\theta=\pi/4$                            & $\varphi=0$, $\theta=\pi/4$ \\
TPX04 & $\varphi\in[0,\pi/4]$, $\theta\in[\pi/3,\pi/2]$        & $\varphi=\pi/6$, $\theta\in[\pi/3,\pi/2]$ \\
TPX05 & $\varphi\in[\pi/2,3\pi/4]$, $\theta\approx\pi/4$       & $\varphi\approx3\pi/4$, $\theta=\pi/4$ \\
\bottomrule
\end{tabular}
\end{minipage}
\end{table}

Analysis methods shown in this section represent the very first attempt in
tracking and particle recognition applied to MoEDAL data. With the arrival of
Timepix3 technology, these techniques were further developed to benefit from
simultaneous ToA+ToT measurement in each pixel, improved time
resolution and data-driven single particle tracking capability~\cite{bergmann20173d,bergmann2021sissa}.

\subsection{Energy loss analysis}\label{sec5sub4}

The mean energy loss $\langle \mathrm{d}E/\mathrm{d}x\rangle$ along each fitted
trajectory is calculated as described in Sect.~\ref{sec5sub1}, by sampling
128~uniform segments and averaging. Under the assumption that the incident
particles are not stopped in the silicon, the per-track averages are aggregated
into the distributions plotted in Fig.~\ref{fig10b},~\ref{fig10d}.

All observed distributions exhibit a monotonically decreasing trend, indicating that
the majority of analyzed particles deposit low energies and that high-deposition
events (reaching up to~$\approx\qty{0.5}{\GeV\per\cm}$) are relatively rare. Local structure visible at low~$\langle \mathrm{d}E/\mathrm{d}x\rangle$ in some
detectors suggests likely existence of dominant species-momentum populations,
which motivates further work toward explicit inference and a more refined
classification.

\begin{figure}[h]
\centering
\begin{minipage}[t]{0.62\textwidth}\centering
\includegraphics[width=\linewidth]{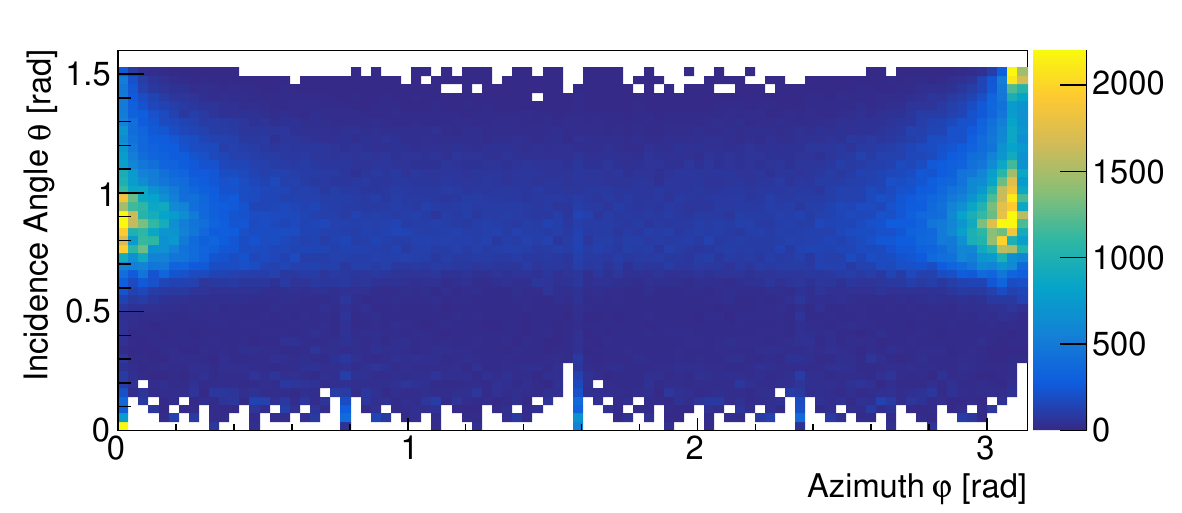}
\\\subfiglabel[TPX03, fitted angles]{fig10a}
\end{minipage}\hfill
\begin{minipage}[t]{0.36\textwidth}\centering
\includegraphics[width=\linewidth]{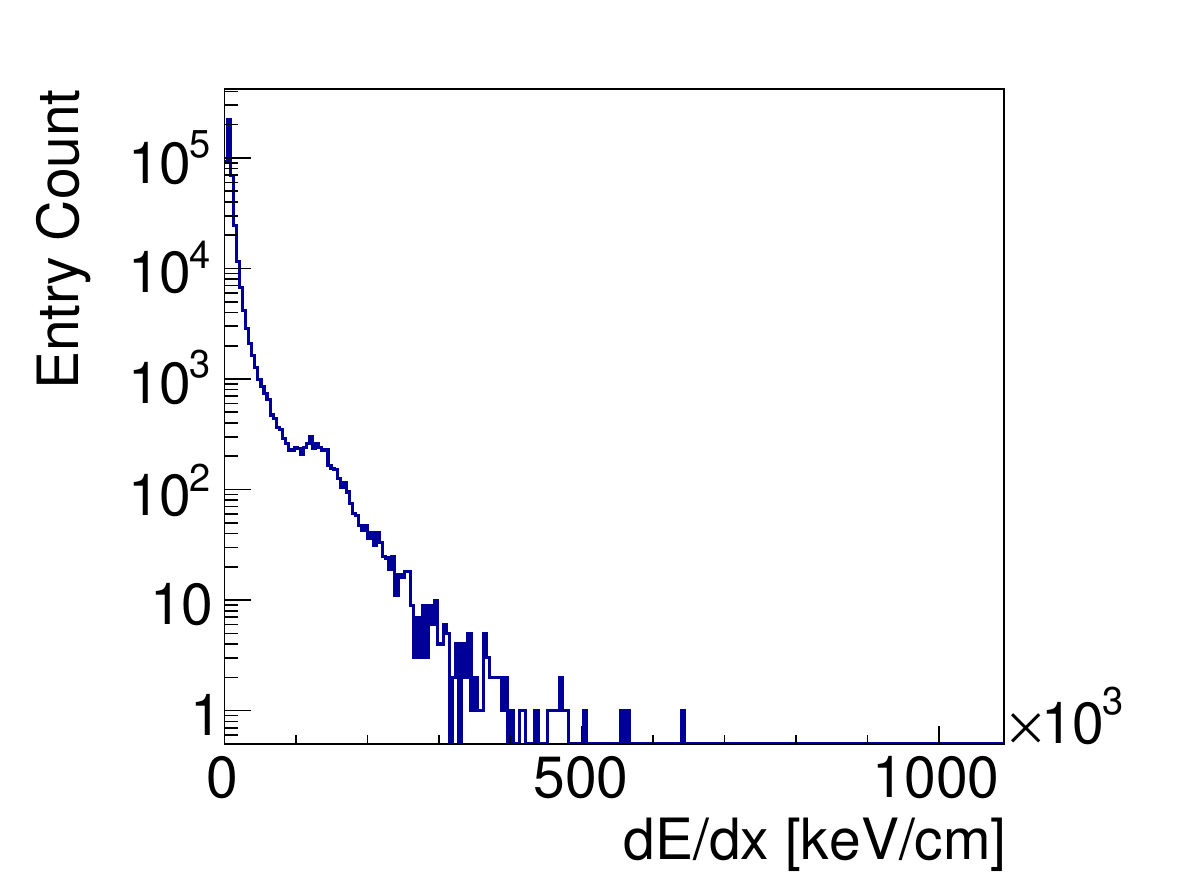}
\\\subfiglabel[TPX03, energy loss]{fig10b}
\end{minipage}
\\
\begin{minipage}[t]{0.62\textwidth}\centering
\includegraphics[width=\linewidth]{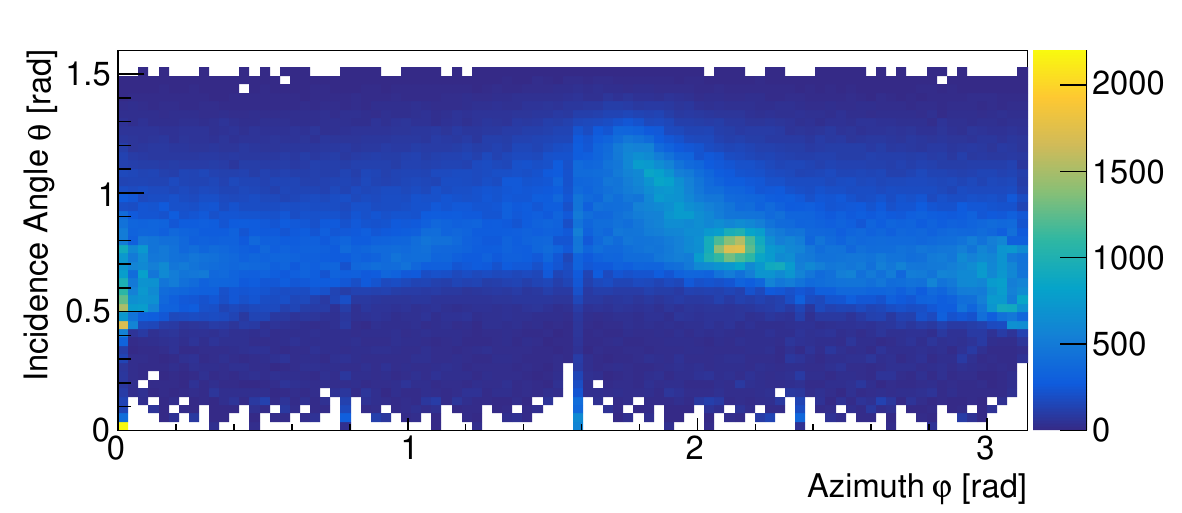}
\\\subfiglabel[TPX05, fitted angles]{fig10c}
\end{minipage}\hfill
\begin{minipage}[t]{0.36\textwidth}\centering
\includegraphics[width=\linewidth]{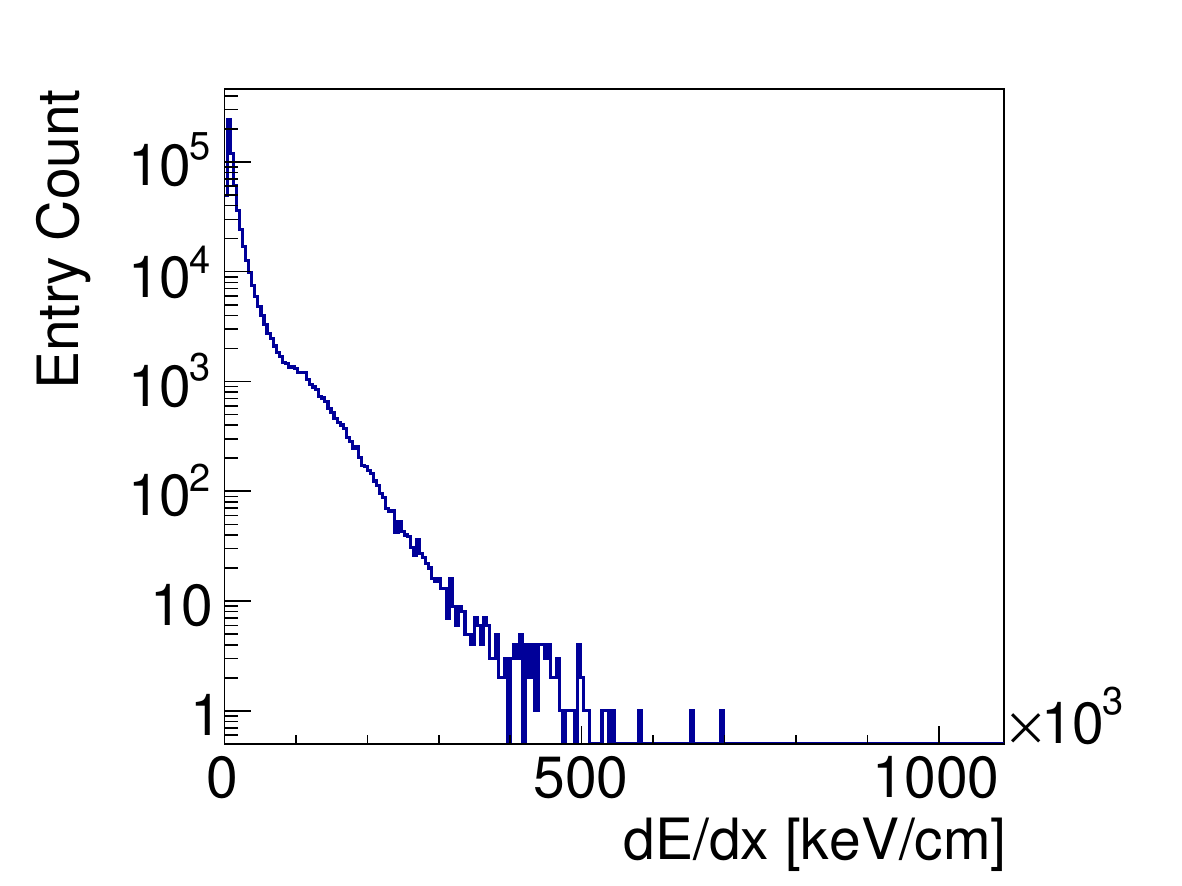}
\\\subfiglabel[TPX05, energy loss]{fig10d}
\end{minipage}
\caption{Aggregated results of 3D~particle trajectory reconstruction, presented
for detectors TPX03~(upper row) and TPX05~(lower row). On the left,
two-dimensional histograms show the distribution of fitted spatial angles~$\varphi$
and~$\theta$, exhibiting localized peaks that represent point sources of
interacting particles (color scale describes event count). On the right, mean
energy loss~$\langle \mathrm{d}E/\mathrm{d}x\rangle$ is plotted in a
one-dimensional frequency distribution, indicating an ample supply of particles
in the range from~\qtyrange{50}{500}{\MeV\per\cm}, relevant to~NTDs. Reproduced from~\cite{Manek2018_CTU}.}\label{fig10}
\end{figure}

\section{Conclusions}\label{sec6}

In this work, the main components of mixed radiation fields present in the
MoEDAL experimental environment were quantified using data recorded by the
Timepix~detector network. Hadronic interactions induced in the silicon
sensors by protons, energetic ions, pions, and charged spallation products were
analyzed using particle track pattern recognition techniques. Given the
essentially \qty[quantity-product={}]{100}{\percent} detection efficiency of
charged ionizing particles in silicon sensors, reliable fluence estimates were
obtained for individual particle groups. For thermal and fast neutrons, absolute
fluences were measured with a dedicated detector on the basis of earlier
calibrations performed with reference neutron sources.

This enabled the primary objective of this study to be achieved: determining the
fluences of individual Standard Model~(SM) particle components that contribute
to the experimental background of the~NTDs. Timepix detects several of the
HIP~species mentioned above. However, in the MoEDAL~NTD~system, these particles
do not generally constitute a physics background that can mimic the signature of
a candidate~HIP associated with new physics~\cite{acharya2014physics}.

Beyond radiation field monitoring, the presented analysis of Timepix data
applied the methodologies and particle-identification techniques described in
Refs.~\cite{bergmann2019characterization,bergmann2021sissa}. Its results
also indicate the feasibility of identifying specific and atypical track
topologies associated with exotic~HIPs. Overall, the demonstrated performance of
Timepix~detectors shows that high‑resolution pixel tracking can significantly
enhance discrimination of primary~HIPs from backgrounds, including neutrons.
This motivates further exploitation of Timepix‑based tracking techniques to
improve SM~background suppression and sensitivity in future searches for
magnetic monopoles and other~HIPs, as well as for broader studies of particle
production in the MoEDAL environment.

\backmatter

\bmhead{Acknowledgements}

The authors gratefully acknowledge the support and cooperation of the Medipix2
and Medipix3~collaborations for instrumentation and ASIC support. In particular,
they thank Michael~Campbell, Erik~Heijne, Xavier~Llopart, and Rafael~Ballabriga,
as well as many colleagues from the MoEDAL~collaboration, for valuable
discussions and guidance on the optimal use of the Timepix detector network
within the MoEDAL experiment.

This work was supported by NSERC~(Canada), the Ministry of Education, Youth and
Sports of the Czech~Republic (project no.~LM2023040), and by the European~Union
under the ROBOPROX project (reg.~no.~CZ.02.01.01/00/22\_008/0004590). The
authors~B.B. and~P.M. acknowledge funding from the Czech Science Foundation
under Registration Number~GM23-04869M.

\bmhead{Dedication}

The authors dedicate this paper to the memory of Prof.~Claude~Leroy, a
valued colleague whose contributions to the experiment and to this work were
greatly appreciated. He will be deeply missed.

\bmhead{Data availability statement}

The datasets generated and analyzed in the presented analysis are available to
readers upon request.


\bmhead{Publication record}

Eur. Phys. J. Spec. Top. (2026). Received 30~April 2026 / Accepted 12~August
2026 / Published online 16~September 2026.
\href{https://doi.org/10.1140/epjs/s11734-026-02586-3}{\nolinkurl{https://doi.org/10.1140/epjs/s11734-026-02586-3}}.
\textcopyright{} The Author(s) 2026.

\bmhead{Open Access}

This article is licensed under a Creative Commons Attribution 4.0 International
License, which permits use, sharing, adaptation, distribution and reproduction
in any medium or format, as long as you give appropriate credit to the original
author(s) and the source, provide a link to the Creative Commons licence, and
indicate if changes were made. The images or other third party material in this
article are included in the article's Creative Commons licence, unless indicated
otherwise in a credit line to the material. If material is not included in the
article's Creative Commons licence and your intended use is not permitted by
statutory regulation or exceeds the permitted use, you will need to obtain
permission directly from the copyright holder. To view a copy of this licence,
visit \href{http://creativecommons.org/licenses/by/4.0/}{\nolinkurl{http://creativecommons.org/licenses/by/4.0/}}.

\bibliography{sn-bibliography}

\end{document}